\documentclass[a4paper,11pt]{article}
\usepackage{jheppub} 

\usepackage{graphicx} 
\usepackage{ytableau} 
\usepackage{cancel}
\usepackage{amsmath}
\usepackage{amssymb}
\usepackage{amsthm}
\usepackage{precomputed_pairs}
\usepackage{calc}
\usepackage{xcolor,array}
\usepackage{mathtools}
\usepackage{tikz}
\usepackage{tikz-cd}
\usepackage{rotating}

\definecolor{lightblue}{RGB}{108,175,230}
\definecolor{lightgreen}{RGB}{130, 247, 124}
\definecolor{lightgray}{RGB}{156, 156, 156}
\definecolor{purple}{RGB}{212, 145, 237}
\theoremstyle{definition}
\newtheorem{definition}{Definition}

\makeatletter
\newcommand{\dalembertian}{\mathop{\mathpalette\dalembertian@\relax}}
\newcommand{\dalembertian@}[2]{
  \begingroup
  \sbox\z@{$\m@th#1\square$}
  \dimen0=\fontdimen8
    \ifx#1\displaystyle\textfont\else
    \ifx#1\textstyle\textfont\else
    \ifx#1\scriptstyle\scriptfont\else
    \scriptscriptfont\fi\fi\fi3
  \makebox[\wd\z@]{
    \hbox to \ht\z@{
      \vrule width \dimen0
      \kern-\dimen0
      \vbox to \ht\z@{
        \hrule height \dimen0 width \ht\z@
        \vss
        \hrule height 2\dimen0
      }
      \kern-2.5\dimen0
      \vrule width 2.5\dimen0
    }
  }
  \endgroup
}
\makeatother
\usepackage{soul}

\newcommand{\delboxbar}[1]{
\none[\text{\begin{tikzpicture}[baseline=(X.base)]
      \node (X) {\raisebox{-0.35ex}{$\overline{#1}$}}; 
      \begin{scope}
        \draw[dashed] 
          ([xshift=-0.016em, yshift=-0.3ex] X.north west) 
          rectangle 
          ([xshift=0.07em, yshift=0.31ex] X.south east); 
      \end{scope}
  \end{tikzpicture}}]
}

\newcommand{\delbox}[1]{
\none[\text{\begin{tikzpicture}[baseline=(X.base)]
      \node (X) {\raisebox{-0.23225ex}{$#1$}}; 
      \begin{scope}
        \draw[dashed] 
          ([xshift=-0.016em, yshift=-0.05ex] X.north west) 
          rectangle 
          ([xshift=0.07em, yshift=0.1ex] X.south east); 
      \end{scope}
  \end{tikzpicture}}]
}

\newcommand{\greendelbox}[1]{
\none[\text{\begin{tikzpicture}[baseline=(X.base)]
    \node[inner sep=0pt] (X) at (0,0) {};
    \fill[lightgreen!100] (-0.645em,0.645em) rectangle (0.64em,-0.65em);
    \draw[dashed] (-0.645em,0.645em) rectangle (0.645em,-0.645em);
    \node at (0,0) {\raisebox{-0.23225ex}{$#1$}};
  \end{tikzpicture}}]
}

\newcommand{\colourdelbox}[2]{
\none[\text{\begin{tikzpicture}[baseline=(X.base)]
    \node[inner sep=0pt] (X) at (0,0) {};
    \fill[#1!100] (-0.645em,0.645em) rectangle (0.64em,-0.65em);
    \draw[dashed] (-0.645em,0.645em) rectangle (0.645em,-0.645em);
    \node at (0,0) {\raisebox{-0.23225ex}{$#2$}};
  \end{tikzpicture}}]
}

\newcommand{\canceldiag}[1]{
  \tikz[baseline=(X.base)]{
    \node (X) {\ensuremath{#1}};
    \draw[black, line width=0.4pt] (X.south west) -- (X.north east);
  }
}

\newcommand{\cancelbox}[1]{
*(#1)\text{
  \begin{tikzpicture}[baseline=(current bounding box.center)]
    \draw[line width=0.4pt] (0,0.6em) -- (0,-1.2em);
    \node at (0,0) {\empty};
  \end{tikzpicture}
}
}

\newcommand{\splitbox}[2]{
  \mathrel{
    \begin{tikzpicture}[baseline=(diag.base)]
      \node[inner sep=0pt, minimum size=1.6ex] (diag) at (0,0) {};
      \draw[line width=0.3pt] (diag.center) -- ++(0.7ex,0.7ex);
      \draw[line width=0.3pt] (diag.center) -- ++(-0.7ex,-0.7ex);
      \node at ([shift={(-0.6ex,0.7ex)}]diag.center) {\scriptsize {$\overline{#1}$}};
      \node at ([shift={(0.5ex,-0.6ex)}]diag.center) {\scriptsize $#2$};
    \end{tikzpicture}
  }
}

\newcommand{\delsplitbox}[2]{
  \none[{
    \mathrel{
      \begin{tikzpicture}[baseline=(X.base)]
        \node (X) {
          \begin{tikzpicture}[baseline=(diag.base)]
            \node[inner sep=0pt, minimum size=0ex] (diag) at (0,0) {};
            \draw[line width=0.3pt] (diag.center) -- ++(0.7ex,0.7ex);
            \draw[line width=0.3pt] (diag.center) -- ++(-0.7ex,-0.7ex);
            \node at ([shift={(-0.6ex,0.7ex)}]diag.center) {\scriptsize {$\overline{#1}$}};
            \node at ([shift={(0.5ex,-0.6ex)}]diag.center) {\scriptsize $#2$};
          \end{tikzpicture}
        };
        \draw[dashed] 
          ([xshift=0.47em, yshift=-1.39ex] X.north west)
          rectangle 
          ([xshift=-0.475em, yshift=1.39ex] X.south east);
      \end{tikzpicture}
    }
  }]
}

\newcommand{\SU}{\mathrm{SU}}
\newcommand{\U}{\mathrm{U}}
\newcommand{\GL}{\mathrm{GL}}
\newcommand{\SL}{\mathrm{SL}}
\newcommand{\appref}[1]{Appendix~\ref{#1}}
\newcommand{\secref}[1]{Section~\ref{#1}}
\newcommand{\Eqref}[1]{eq.~(\ref{#1})}

\usepackage{multicol}
\usepackage{color}
\newcommand{\sm}{\raisebox{0.2ex}{\scalebox{0.65}{$-$}}}

\newtheoremstyle{algorithm}
  {\topsep}
  {\topsep}
  {}
  {0pt}
  {\bfseries}
  {}
  { }
  {\thmname{\bfseries #1}\thmnumber{\ \bfseries #2}.\thmnote{ \bfseries\boldmath #3}}

\theoremstyle{algorithm}

\newtheorem{theorem}{Algorithm}

\definecolor{forestgreen}{RGB}{0, 135, 52}

\arxivnumber{} 

\title{\boldmath An $N$-independent tensor decomposition for $\SU(N)$}

\author[a]{Stefan Keppeler,}
\author[b]{Malin Sjodahl}
\author[c,d]{and Bernanda Telalovic}
\affiliation[a]{Universit\"at T\"ubingen, Mathematisches Institut, Auf der Morgenstelle 10,
D-72076 T\"ubingen, Germany}
\affiliation[b]{Department of Physics, Lund University, Box 118, SE-221 00 Lund, Sweden}
\affiliation[c]{Niels Bohr International Academy, Københavns Universitet, Blegdamsvej 17, DK-2100 Copenhagen, Denmark}
\affiliation[d]{Institute of Physics, NAWI Graz, University of Graz, Universit\"atsplatz 5, A-8010 Graz, Austria}

\date{January 2025}

\emailAdd{stefan.keppeler@uni-tuebingen.de}
\emailAdd{malin.sjodahl@fysik.lu.se}
\emailAdd{bernanda.telalovic@nbi.ku.dk}

\abstract{To facilitate a simultaneous treatment of an arbitrary number of colors in representation theory-based descriptions of QCD color structure, we derive an $N$-independent reduction of $\SU(N)$ tensor products. To this end, we label each irreducible representation by a pair of Young diagrams, with parts acting on quarks and antiquarks. By combining this with a column-wise multiplication of Young diagrams, we generalize the Littlewood--Richardson rule for the product of two Young diagrams to the product of two Young diagram pairs, achieving a general-$N$ decomposition.}

\begin{document}
\maketitle
\flushbottom

\section{Introduction}
\ytableausetup{boxsize=1.25em}
A major computational bottleneck in the calculation of scattering amplitudes for many partons is the color structure of the strong force.
The most common approaches for treating the $\SU(N)$ color structure of scattering amplitudes are to use trace bases \cite{Paton:1969je, Dittner:1972hm, Cvitanovic:1976am,Berends:1987cv, Mangano:1987xk, Mangano:1988kk, Kosower:1988kh, Nagy:2007ty, Sjodahl:2009wx, Alwall:2011uj, Sjodahl:2012nk,Sjodahl:2014opa, Platzer:2012np,Platzer:2018pmd,Frederix:2021wdv} or color flow bases \cite{tHooft:1973alw,Kanaki:2000ms,Maltoni:2002mq, Kilian:2012pz, Platzer:2013fha,AngelesMartinez:2018cfz,DeAngelis:2020rvq,Platzer:2020lbr,Frederix:2021wdv}. For finite $N$, these spanning sets are in general non-orthogonal and overcomplete, with the number of basis vectors scaling approximately factorially with the number of gluons plus quark--antiquark pairs \cite{Sjodahl:2015qoa}. 
When dealing with a high number of partons, this implies that the treatment of the color space becomes so time-consuming that one may choose to resort to an expansion in the number of colors in the large-$N$ limit \cite{Frederix:2021wdv}, or to a sampling of color configurations \cite{Platzer:2013fha,Isaacson:2018zdi,AngelesMartinez:2018cfz,DeAngelis:2020rvq,Platzer:2020lbr, Forshaw:2025bmo, Forshaw:2025fif}. 

Alternatively, one may want to employ orthogonal multiplet bases, where the basis vectors are characterized by irreducible representations, which in turn are labeled by Young diagrams. 
The method of using Young diagrams to characterize irreducible representations goes back to the early 1900s \cite{Young:collected}. For the special unitary group $\SU(N)$ multiplication of Young diagrams via the Littlewood--Richardson rule \cite{Littlewood:1934} directly gives the irreducible representations arising in the Clebsch--Gordan decomposition of $\SU(N)$ product representations. For example, for $\SU(3)$ and the gluon octet of the strong force one finds the familiar decomposition  $8 \otimes 8=1 \oplus 8 \oplus 8 \oplus 10 \oplus \overline{10} \oplus 27$.

In quantum chromodynamics (QCD), quarks, antiquarks and gluons transform under the fundamental, antifundamental, and adjoint representations of $\SU(3)$, labeled by the Young diagrams \scalebox{0.6}{$\ydiagram{1}$}, \raisebox{1ex}{\scalebox{0.6}{$\ydiagram{1,1}$}}, and \raisebox{1ex}{\scalebox{0.6}{$\ydiagram{2,1}$}}, respectively. For four colors, quarks, antiquarks, and gluons would instead transform under the $\SU(4)$ representations labeled by the Young diagrams \scalebox{0.6}{$\ydiagram{1}$}, \raisebox{2ex}{\scalebox{0.6}{$\ydiagram{1,1,1}$}}, and \raisebox{2ex}{\scalebox{0.6}{$\ydiagram{2,1,1}$}}, respectively. Comparing the same calculation for different $\SU(N)$ gauge groups thus requires working with a different set of Young diagrams for each $N$. 
However, it is often enlightening to be able to retain the full $N$-dependence, both to study the analytic structure, and to be able to take the leading-$N$
limit.

Alternatively, irreducible representations of $\SU(N)$ can be labeled by \textit{pairs} of Young diagrams \cite{King:1970, Sjodahl:2018cca}. For example, in our notation for Young diagram pairs (to be introduced in \secref{sec:pair notation}), quarks, antiquarks and gluons, for every $N$, transform under representations labeled by \scalebox{0.6}{$\ydiagram{1}$}, \scalebox{0.6}{$\ydiagram[\bullet]{1}$}, and \raisebox{1ex}{\scalebox{0.6}{$\begin{ytableau}\none&\empty\\\bullet\end{ytableau}$}}, respectively. 
Using Young diagram pairs, we develop and prove rules for decomposing the tensor product of two $\SU(N)$ representations,
simultaneously valid for any $N$.
We refer to such a calculation as an $N$-independent or \textit{general-$N$} calculation.

For treating the $\SU(N)$ color structure of QCD, it is, of course, not sufficient to know the decomposition of product representations into direct sums of irreducible representations. Instead, for a representation theory-based approach, one has to construct multiplet bases consisting of Hermitian projection operators to irreducible representations and transition operators between equivalent representations. This is a complicated task, first addressed for a low number of partons \cite{MacFarlane:1968vc,Butera:1979na,Kyrieleis:2005dt,Dokshitzer:2005ig,Sjodahl:2008fz,Beneke:2009rj}, and only later worked out in generality, first for standard model representations, \cite{Keppeler:2012ih, Du:2015apa,Sjodahl:2015qoa,Keppeler:2013yla,Alcock-Zeilinger:2016bss, Alcock-Zeilinger:2016sxc,Alcock-Zeilinger:2016cva,Sjodahl:2018cca}, and for exotic representations~\cite{Ohl:2024} only recently~\cite{Sjodahl:2024fqn}. Novel approaches for calculating Wigner~6j coefficients may in principle allow for circumventing the explicit basis construction~\cite{Alcock-Zeilinger:2022hrk, Keppeler:2023msu}, but also in this case, it is essential to have a tensor reduction procedure simultaneously valid for all $N$, in order to dissect the $N$-dependence. 

Combining or connecting representation theory-based fixed-$N$ methods 
and large-$N$ expansions, requires general-$N$ calculations in the sense defined above. In particular, basis vectors for states built from quarks, antiquarks and gluons have to be labeled by Young diagram pairs instead of by single Young diagrams (which would be sufficient for states built from only quarks or only antiquarks), and tensor products have to be performed keeping this labeling scheme throughout the calculation. 

King, who in \cite{King:1970} introduced the labeling in terms of Young diagram pairs, in the same work also gives an algorithm for decomposing tensor products of representations labeled by Young diagram pairs into direct sums of irreducible representations also labeled by Young diagram pairs. King's method, however, leads to an overcounting of terms, i.e., in intermediate steps additional terms are generated, which later have to be subtracted. This makes the algorithm costly to implement. Our method (Algorithm~\ref{alg:pair_multiplication} in \secref{sec:pair multiplication}) avoids this overcounting and constructs all terms of the decomposition in a manner similar to the Littlewood--Richardson rule. 

To achieve the general-$N$ description, we first, in \secref{sec:notation}, introduce our notation for ordinary and barred Young diagrams, and explain how to interpret them for fixed~$N$. We then, in \secref{sec:Column multiplication}, cast the standard Littlewood--Richardson rule for multiplying Young diagrams --- which operates at the level of rows --- into a column-based version. In \secref{sec:pair notation} we {introduce} the general-$N$ labeling of irreducible representations in terms of Young diagram pairs. In \secref{sec:algorithms_for_special_cases} we discuss multiplication of a Young diagram pair by an ordinary Young diagram or by a barred Young diagram, and in \secref{sec:pair multiplication} we combine these special cases into our main result, Algorithm~\ref{alg:pair_multiplication}, for multiplying two Young diagram pairs. We formulate our results for $\SU(N)$ and explain in \appref{app:GLN} in which sense they are also valid for groups such as $\U(N)$ or $\GL(N)$. We provide conclusions in \secref{sec:conclusion}.

\section{Young diagrams and barred Young diagrams labeling representations of {\boldmath\texorpdfstring{$\SU(N)$}{SU(N)}}}
\label{sec:notation}
\ytableausetup{boxsize=1.25em}

Irreducible representations of $\SU(N)$ can be labeled by Young diagrams. The irreducible representation labeled by a diagram with $n$ boxes is carried by a subspace of the tensor product of $n$ copies of the carrier space of the fundamental or defining representation of $\SU(N)$, i.e., the representation in terms of unitary $N\times N$ matrices with unit determinant. 
The subspace corresponding to a given Young diagram is the image of a Young operator, which is a product of symmetrizers and antisymmetrizers over subsets of the $n$ factors, see e.g., \cite{Tung:1985,Cvitanovic:2008}. 
Columns with more than $N$ boxes correspond to antisymmetrization of more than $N$ factors and thus the image of Young operators corresponding to Young diagrams with more than $N$ rows is empty. 
A column with exactly $N$ boxes contributes a factor $\det g=1$ to the representation matrix of the group element $g\in\SU(N)$ and thus columns of length $N$ can be omitted. For details of all this see e.g., \cite{Littlewood:1950,Hamermesh:1962,Lichtenberg:1978,Tung:1985,Jones:1990,Cvitanovic:2008}.

We denote Young diagrams by Greek letters and identify them with their
non-increasing sequence of rows lengths
$\lambda=(\lambda_1,\lambda_2,...,\lambda_l)$, e.g.
\begin{equation}\ytableausetup{boxsize=1.25em, centerboxes,centertableaux}
    (4,2,1)
    =\scalebox{0.7}{\ydiagram{4,2,1}}
    \overset{N=3}{=}\scalebox{0.7}{$\ydiagram{3,1}$}\ ,
\end{equation}
where the last equality illustrates the rule that columns of length $N$ can be omitted.

The fundamental representation, labeled by a Young diagram consisting of a single box, \raisebox{0.5ex}{\scalebox{0.6}{$\ydiagram{1}$}}, is carried by an $N$-dimensional vector space. The dual space carries another $N$-dimensional representation, which 
for $\SU(N)$ is equivalent\footnote{Recall that two representations are called equivalent if the representation matrices of one representation can be obtained by a similarity transformation from the representation matrices of the other representation.} to the complex conjugate of the fundamental representation, i.e., the antifundamental representation, which we denote by \raisebox{0.5ex}{\scalebox{0.6}{${\overline{\ydiagram{1}}}$}}. 
One can show that this representation is equivalent to the representation labeled by a Young diagram consisting of a single column of length $N-1$, see e.g., \cite{Lichtenberg:1978}. This observation generalizes as described in the following paragraph. 

We associate with each Young diagram a barred diagram which we obtain from the original diagram by rotating it by $180^\circ$ and then removing its boxes from a rectangular diagram of height $N$ and same width as the original diagram, e.g.,
\begin{equation}
  \begin{split}\ytableausetup{boxsize=1.25em}
    \scalebox{0.7}{$\ydiagram{3,1}$}
    \xrightarrow{\text{rotate}}
         \scalebox{0.7}{$\ydiagram[*(white)\bullet]{2+1,3}$} \xrightarrow{N{=}4}  \scalebox{0.7}{$\begin{ytableau}
         \none[\scriptstyle N] & \none[\scriptstyle N] & \none[\scriptstyle N]\\
             *(lightgray)\empty & *(lightgray)\empty & *(lightgray)\empty \\
             *(lightgray)\empty & *(lightgray)\empty & *(lightgray)\empty \\
             *(lightgray)\empty & *(lightgray)\empty & *(lightgray)\bullet \\
             *(lightgray)\bullet & *(lightgray)\bullet & *(lightgray)\bullet 
         \end{ytableau}$}
         \xrightarrow{\text{remove }{\scalebox{0.7}{$\onebox{\bullet}$}}}  \scalebox{0.7}{$\begin{ytableau}
         \none[\text{\begin{rotate}{45}$\scriptstyle N\sm 1$\end{rotate}}] & \none[\text{\begin{rotate}{45}$\scriptstyle N\sm 1$\end{rotate}}] & \none[\text{\begin{rotate}{45}$\scriptstyle N\sm 2$\end{rotate}}]\\
             \empty & \empty & \empty \\
             \empty & \empty & \empty \\
             \empty & \empty  \\
           \end{ytableau}$}
         \ \stackrel{N{=}4}{=} \ \overline{\scalebox{0.7}{$\ydiagram{3,1}$}}\ ,
    \end{split}
    \label{eq:cut out}
\end{equation}
where we have filled boxes of the rotated diagram and boxes to be removed with bullets. Note that the shape of the resulting diagram depends on $N$. In the example we have displayed the actual diagram for $N=4$ and have indicated the column lengths of the diagram for arbitrary $N$ above the columns. One can show that for $\SU(N)$ the representation labeled by the barred diagram is equivalent to the complex conjugate of the representation labeled by the original diagram, see e.g., \cite{Lichtenberg:1978}.

In QCD, quarks transform under the fundamental representation and antiquarks transform under the antifundamental representation. In the light of the discussion above, the identity
\begin{equation}
  \overline{\scalebox{0.7}{$\ydiagram{3,1}$}}
  \ \stackrel{N{=}3}{=} \
  \scalebox{0.7}{$\ydiagram{3,2}$}
\end{equation}
means that there exists a multiplet of four-antiquark states and a multiplet of five-quark states which transform under equivalent representations. 
The barred diagrams allow us to refer to multiplets of antiquarks in an $N$-independent way. 

In the following we find it convenient to denote barred diagrams by rotated diagrams with boxes filled with bullets, e.g.,
\begin{equation}\label{eq:bullet-diagrams}
  \overline{\scalebox{0.7}{$\ydiagram{3,1}$}}\,
  =\scalebox{0.7}{$\ydiagram[*(white)\bullet]{2+1,3}$}\, .  
\end{equation}
This notation visually hints at the shape of the ordinary diagram labeling the equivalent representation, since this ordinary diagram is obtained by removing the boxes filled with bullets from a rectangular diagram, see \Eqref{eq:cut out}. The full power of this notation will only become clear when we introduce Young diagram pairs in \secref{sec:pair notation}. {In the following we will refer to the diagrams on both sides of \Eqref{eq:bullet-diagrams} as barred (Young) diagrams, and we will also refer to \raisebox{0.5ex}{\scalebox{0.6}{$\ydiagram[\bullet]{1}$}} as a barred box.}

\section{Column-wise multiplication of Young diagrams}
\label{sec:Column multiplication}

Tensor products of irreducible representations of $\SU(N)$ are
reducible to direct sums of irreducible representations,
\begin{equation}\label{eq:reduce-lambda-times-mu}
  \lambda \otimes \mu = \bigoplus_\nu c_{\lambda,\mu}^\nu \nu \, ,
\end{equation}
where the multiplicities $c_{\lambda,\mu}^\nu$ are known as
Littlewood--Richardson numbers~\cite{Littlewood:1934,Fulton:1997,Sagan:2000} 
and obviously
satisfy $c_{\lambda,\mu}^\nu=c_{\mu,\lambda}^\nu$. These numbers can
be determined by the algorithm below, which is explained in most
textbooks on group and representation theory, see e.g.,
\cite{Littlewood:1950,Hamermesh:1962,Tung:1985,Jones:1990,Cvitanovic:2008}; 
proofs are, e.g., given
in \cite{Fulton:1997,Sagan:2000}. 
As the algorithm is treating one row at a time,
we refer to it as 
row-wise multiplication of Young diagrams, or row
multiplication for short. Below, we give a version of the
algorithm defining a multiplication of Young diagrams
in their own right. The restriction to any finite $N$ is simply obtained by 
omitting Young diagrams containing columns of length
$>N$.

In the standard algorithm for multiplication of Young diagrams, as well as in later algorithms, we need to refer to the admissibility of
a sequence:
\begin{definition}\label{def:admissible}
A sequence is called \textit{admissible} if for all $j<k$, 
        both appearing at least once, the sequence, when
        terminated at any point, contains at least as many numbers
        $j$ as it contains numbers $k$. 
\end{definition}
For instance, the sequence 1, 2, 1, 2, 3, 3 is admissible, whereas 1, 2, 1, 3, 3, 2 is not since, when terminated after the 5th term, it contains more 3s than 2s. 
In the literature, admissible sequences are also known as lattice words \cite{Fulton:1997} or lattice permutations \cite{Sagan:2000}. 
With the notion of admissibility at hand, we can state the algorithm for multiplying Young diagrams, known as the Littlewood--Richardson rule.

\begin{theorem}[Row-wise multiplication of Young diagrams]\label{alg:row_multiplication} \ \\
The
multiplicities
$c_{\lambda,\mu}^\nu$ showing up in the
reduction of $\lambda\otimes\mu$ can be obtained as follows:
\begin{enumerate}
    \item Label each box in $\mu$ by the row in which it appears.
    \item Starting from row one, for each row $j$, in $\mu$, append all boxes,
    each labeled $j$, to $\lambda$ in such a way that the resulting
    semitableau\footnote{Any Young diagram that is only partially filled
    with numbers we refer to as a semitableau.}
    \begin{enumerate}
        \item has the shape of a Young diagram,
        \item does not have any two boxes labeled $j$ appearing in the
        same column,
        \item has a row-sequence of numbers, read
        right-to-left and then top-to-bottom that is
        admissible.
\end{enumerate}
    \item After exhausting the rows of $\mu$, the multiplicity 
    $c_{\lambda,\mu}^\nu$ is given by the number of semitableaux of shape $\nu$. 
\end{enumerate}
\end{theorem}

We recall that,
when reducing an $\SU(N)$ product representation  $\lambda\otimes\mu$ for a particular value of $N$, the sum on the r.h.s.\ of \Eqref{eq:reduce-lambda-times-mu} extends only over those $\nu$
that result from row multiplication \emph{and} have at most $N$
rows. In practice, for fixed $N$, it is more convenient to cross out tableaux with more than $N$ rows as they are encountered,
but applying the finite-$N$ cross-out later is equivalent and will provide 
a more direct path to column multiplication. Moreover, columns of length $N$ may be omitted from the diagrams
$\nu$, after having determined the multiplicity.

We now introduce the transpose $\mathcal{T}(\lambda)$ of a Young
diagram $\lambda$ (and similarly for tableaux and semitableaux) which
turns rows into columns and vice versa, e.g.,
\begin{equation}\ytableausetup{centerboxes}
    \mathcal{T}\left(\scalebox{0.7}{$\ytableaushort{\empty\empty\empty,\empty}$}\right)
    = \scalebox{0.7}{$\ytableaushort{\empty\empty,\empty,\empty}$}\, \ , \, \quad
    \mathcal{T}\left(\scalebox{0.7}{$\ytableaushort{\empty12,34}$}\right)
    = \scalebox{0.7}{$\ytableaushort{\empty3,14,2}$} \;.
\end{equation}
Obviously, $\mathcal{T}\circ\mathcal{T}$ is the identity if we again
view Young diagrams as entities in their own right, i.e., in
particular without restricting to a maximal number of rows. One can
show \cite[Section~5.1, Corollary~2]{Fulton:1997} 
that the
Littlewood--Richardson numbers satisfy
\begin{equation}\label{eq:transposed-LR-numbers}
  c_{\mathcal{T}(\lambda),\mathcal{T}(\mu)}^{\mathcal{T}(\nu)}
  = c_{\lambda,\mu}^\nu. 
\end{equation}
This property allows us to prove (in \appref{app:Column-wise proof}) the following alternative algorithm for multiplying Young diagrams, which we call column-wise
multiplication of Young diagrams, or column multiplication for short.

\begin{theorem}[Column-wise multiplication of Young diagrams]\label{alg:column_multiplication} \ \\
The multiplicities $c_{\lambda,\mu}^\nu$ showing up in the
reduction of $\lambda\otimes\mu$ can be obtained as follows:
\begin{enumerate}
    \item Label each box in $\mu$ by the column in which it appears.
    \item Starting from column one, for each column $j$, in $\mu$, append all
    boxes, each labeled $j$, to $\lambda$ in such a way that the resulting
    semitableau
    \begin{enumerate}
        \item has the shape of a Young diagram,
        \item does not have any two boxes labeled $j$ appearing in the
        same \textit{row},
        \item has a column-sequence of numbers, read
        \textit{bottom-to-top} and then \textit{left-to-right} that is admissible (see Definition~\ref{def:admissible}).
\end{enumerate}
    \item After exhausting the columns of $\mu$, the multiplicity
    $c_{\lambda,\mu}^\nu$ is given by the number of semitableaux of shape $\nu$. 
\end{enumerate}
\end{theorem}

We remark that within the proof establishing the equivalence of row-wise and column-wise multiplication of Young diagrams (see \appref{app:Column-wise proof}), we treat Young diagrams as entities in their own right. 
However, when applied to the reduction of a product representation of $\SU(N)$ for a particular value of $N$, one can use either of the two algorithms and immediately discard diagrams (or semitableaux) with more than $N$ rows whenever they appear.

Note that although the Young diagrams $\nu$ and their multiplicities $c_{\lambda,\mu}^\nu$ resulting from column multiplication agree with those obtained by row multiplication, the semitableaux appearing in intermediate steps are in general different for the two algorithms.  

We conclude this section by applying column multiplication to a standard example,
\begin{equation}\label{eq:octet_mult}
\ytableausetup{mathmode, centerboxes,centertableaux}
    \begin{split}
        \scalebox{0.7}{$\ydiagram{2,1}$}
        \,\otimes \, 
        \scalebox{0.7}{$\octet $}
        =& \ 
        \scalebox{0.7}{$\ydiagram{2,1}$}
        \,\otimes \, 
        \scalebox{0.7}{$\ytableaushort{12,1}$}\\
        =& \left(\canceldiag{\scalebox{0.7}{$\ytableaushort{\empty\empty11,\empty}$}}\,\oplus\,
                 \scalebox{0.7}{$\ytableaushort{\empty\empty1,\empty1}$}\,\oplus\,
                 \scalebox{0.7}{$\ytableaushort{\empty\empty1,\empty,1}$}\, \oplus\,
                 \scalebox{0.7}{$\ytableaushort{\empty\empty,\empty1,1}$}\, \oplus\,
                 \scalebox{0.7}{$\ytableaushort{\empty\empty,\empty,1,1}$}\right)\,\otimes\,\scalebox{0.7}{$\ytableaushort{2}$}\\
        =& \ \scalebox{0.7}{$\ytableaushort{\empty\empty12,\empty1}$}\,\oplus\,
           \scalebox{0.7}{$\ytableaushort{\empty\empty1,\empty12}$}\,\oplus\,
           \canceldiag{\scalebox{0.7}{$\ytableaushort{\empty\empty1,\empty1,2}$}}\,\oplus\,
           \scalebox{0.7}{$\ytableaushort{\empty\empty12,\empty,1}$}\,\oplus\,
           \scalebox{0.7}{$\ytableaushort{\empty\empty1,\empty2,1}$}\,\oplus\,
           \canceldiag{\scalebox{0.7}{$\ytableaushort{\empty\empty1,\empty,1,2}$}}\\
        & \text{\small \hspace{8pt} (1,1,2) \hspace{15pt} (1,2,1) \hspace{13pt} (2,1,1) \hspace{21pt} (1,1,2) \hspace{18pt} (1,2,1) \hspace{16pt} (2,1,1)}\\
        &\oplus\,\scalebox{0.7}{$\ytableaushort{\empty\empty2,\empty1,1}$}\,\oplus\,
            \scalebox{0.7}{$\ytableaushort{\empty\empty,\empty1,12}$}\,\oplus\,
            \canceldiag{\scalebox{0.7}{$\ytableaushort{\empty\empty,\empty1,1,2}$}}\,\oplus\,
           \scalebox{0.7}{$\ytableaushort{\empty\empty2,\empty,1,1}$}\,\oplus\,\ 
           \scalebox{0.7}{$\ytableaushort{\empty\empty,\empty2,1,1}$}\,\oplus\,\ 
           \canceldiag{\scalebox{0.7}{$\ytableaushort{\empty\empty,\empty,1,1,2}$}}\\
        & \text{\small \hspace{11pt} (1,1,2) \hspace{9pt} (1,2,1) \hspace{5pt} (2,1,1) \hspace{10pt} (1,1,2) \hspace{12pt} (1,1,2) \hspace{10pt}  (2,1,1)}\\
        =& \ \scalebox{0.7}{$\ydiagram{4,2}$}\,\oplus\, 
        \scalebox{0.7}{$\ydiagram{3,3}$}\,\oplus\,
        \scalebox{0.7}{$\ydiagram{4,1,1}$}\,\oplus\,
        2\ \scalebox{0.7}{$\ydiagram{3,2,1}$}\,\oplus\,
        \scalebox{0.7}{$\ydiagram{2,2,2}$}\,\oplus\,
        \scalebox{0.7}{$\ydiagram{3,1,1,1}$}\,\oplus\,
        \scalebox{0.7}{$\ydiagram{2,2,1,1}$}
         \ ,
    \end{split}
\end{equation}
where the column-sequences are shown below each semitableau in the second-to-last line. Note that here we have also displayed (and crossed out) some terms which are only ruled out by the admissibility criterion for the column sequence. Also note that line 2 of this equation (and similarly in other equations below) has no meaning as an identity for Young diagrams or representations, but is simply a useful illustration of the step-by-step application of Algorithm~\ref{alg:column_multiplication}. Finally, we remark that for $N=3$ we recover the standard result of $8 \otimes 8$.

\section{Young diagram pairs}
\label{sec:pair notation}
We now introduce the objects of our primary interest, namely irreducible representations of $\SU(N)$ within mixed tensor products of fundamental and antifundamental representations. Our aim is to reduce tensor products of such representations simultaneously for all $N$. We label these representations by pairs of Young diagrams (one barred and one ordinary).

By the Young diagram pair $\pair{\rho}{\sigma}$ we denote the irreducible representation carried by that subspace of the carrier space of $\overline{\rho}\otimes\sigma$ on which all contractions between any fundamental and any antifundamental factor vanish \cite{King:1970}, see also \cite[Chapter~13]{Tung:1985}, i.e., no quark-antiquark pair is in a singlet. Equivalently, using the terminology of \cite{Keppeler:2012ih,Sjodahl:2018cca}, the irreducible representation labeled by $\pair{\rho}{\sigma}$ is the unique representation of highest first occurrence within $\overline{\rho}\otimes\sigma$.

Young diagram pairs were first introduced for this purpose by King \cite{King:1970}. King uses a back-to-back notation for Young diagram pairs and calls them composite Young diagrams. The diagrammatic notation for Young diagram pairs differs between texts \cite{King:1970,King_1970,Keppeler:2012ih,GREBlack_1983,CJCummins_1987,Sjodahl:2018cca}, and we here introduce a version adapted to the algorithm for pair multiplication to be presented in \secref{sec:pair multiplication}.

For a given pair $\pair{\rho}{\sigma}$, we extend the Young diagram $\sigma$ in its bottom-left-most corner, by the Young diagram $\overline{\rho}$ for which we use the notation of \Eqref{eq:bullet-diagrams} (i.e., we rotate $\rho$ by $180^\circ$ and fill it with bullets), e.g.,
\begin{equation}\label{eq:pair_notation}
  \pair{\scalebox{0.7}{$\ydiagram{2,1}$}}{\scalebox{0.7}{$\ydiagram{1,1}$}}
  = \left({\scalebox{0.7}{$\begin{ytableau}
        \none &\bullet\\\bullet&\bullet
    \end{ytableau}$}}\, , {\scalebox{0.7}{$\begin{ytableau}
        \empty\\ \empty\\
    \end{ytableau}$}}\right) = {\scalebox{0.7}{$\pairNotationExample$}}\;.
\end{equation}
One advantage of this notation is that the lowest $N$ for which this pair can occur is given by the total number of rows of the diagram pair, i.e., in this case $N_\text{min} = 4$. In this notation, boxes of the barred diagram can never appear in the same row as boxes of the ordinary diagram.

As barred Young diagrams, Young diagram pairs can be mapped to ordinary Young diagrams labeling irreducible representations for fixed $N$.
To this end, we map the barred part of a pair to an ordinary diagram as described in \secref{sec:notation} and attach the ordinary part of the pair on the right. For the Young diagram pair $\pair{(2,1)}{(1,1)}$,  
which has $N_\text{min}=4$, 
see \Eqref{eq:pair_notation}, we obtain
\vspace{-15pt}
\begin{equation}{
    \scalebox{0.7}{$\ydiagram[*(white)\bullet]{0,0,1+1,0+2}*[*(white)]{2+1,2+1}$} \overset{N=4}{=} \scalebox{0.7}{$\begin{ytableau}
      \none\\
          \none[\text{\begin{rotate}{45}$\scriptstyle N\sm 1$\end{rotate}}] & \none[\text{\begin{rotate}{45}$\scriptstyle N\sm 2$\end{rotate}}] & \none[\scriptstyle 2] \\
          *(lightgray)\empty & *(lightgray)\empty & \empty \\
          *(lightgray)\empty & *(lightgray)\empty & \empty\\
          *(lightgray)\empty
          \\
          \none\\\none\\
      \end{ytableau}$}
    \qquad \text{and} \qquad 
    \scalebox{0.7}{$\ydiagram[*(white)\bullet]{0,0,1+1,0+2}*[*(white)]{2+1,2+1}$} \overset{N=5}{=} \scalebox{0.7}{$\begin{ytableau}
      \none\\
          \none[\text{\begin{rotate}{45}$\scriptstyle N\sm 1$\end{rotate}}] & \none[\text{\begin{rotate}{45}$\scriptstyle N\sm 2$\end{rotate}}] & \none[\scriptstyle 2] \\
          *(lightgray)\empty & *(lightgray)\empty & \empty \\
          *(lightgray)\empty & *(lightgray)\empty & \empty\\
          *(lightgray)\empty &*(lightgray)\empty \\
          *(lightgray)\empty\\
          \none\\\none\\
      \end{ytableau}$} \ ,
    }\vspace{-10pt}
\end{equation}
for $N=4$ and $N=5$, respectively. Here we have shaded all boxes of the fixed-$N$ image of the barred part of the pair in gray. 

Note that the number of rows of the fixed-$N$ image of the pair is explicitly $N$-dependent, and determined by the shape of the barred part of the pair, while the number of columns is always fixed independently of $N$. For this reason, it is useful to recast the row form of the standard Littlewood--Richardson multiplication rules into the column form  we have presented, as it can be applied more straightforwardly to the multiplication of a Young diagram pair with both an ordinary and a barred Young diagram, and --- ultimately --- to the product of pairs with pairs. We will proceed by presenting the generalizations in this order in Sections~\ref{sec:algorithms_for_special_cases} and \ref{sec:pair multiplication}.

  \subsection{Dimension of the representation labeled by a pair}
For ordinary Young diagrams, the dimension of the corresponding irreducible $\SU(N)$ representation can be determined as the ratio of an $N$-dependent polynomial and the hook length factor of the Young diagram, see e.g.~\cite[Section~6.1]{Fulton:1991} or \cite[Section~9.4.3]{Cvitanovic:2008}. For diagram pairs, an equivalent relation is derived in \cite[eq.~(45)]{King_1970}. The $N$-dependent dimension, $D_{\pair{\rho}{\sigma}}(N)$, of the $\SU(N)$ representation corresponding to the Young diagram pair $\pair{\rho}{\sigma}$ is given by
\begin{equation}
  D_{\pair{\rho}{\sigma}}(N)
  = \left( \prod_{(i,j)\in\sigma}
  \frac{(N+1-i-j+\sigma_i+\rho_j)}{(1-i-j+\sigma_i+\sigma_j')} \right)
  \left( \prod_{(k,\ell)\in\rho}
  \frac{(N-1+k+l-\sigma_k'-\rho_\ell')}{(1-k-\ell+\rho_k+\rho_\ell')} \right) .
\end{equation}
Here the first product is over all boxes of $\sigma$, with $(i,j)$ denoting the box in row $i$ and column $j$. Similarly, $(k,l) \in \rho$ denotes the box in row $k$ and column $\ell$ of $\rho$. The length of the $i$th row of $\sigma$ is denoted by $\sigma_i$ and the length of its $j$th column by $\sigma_j'$, and similarly for $\rho$.

\section{Multiplying a Young diagram pair by an ordinary Young diagram or by a barred Young diagram}\label{sec:algorithms_for_special_cases}

\subsection{Multiplying a Young diagram pair by an ordinary Young diagram}\label{sec:pair x ordinary}
The extension of column multiplication to the multiplication of a diagram pair by an ordinary Young diagram, i.e., $\pair{\rho}{\sigma}\otimes\mu$, is straightforward, but two aspects deserve special attention. First, boxes from $\mu$ can either extend the diagram $\sigma$, or \textit{remove} boxes from $\rho$ \cite{Sjodahl:2018cca}, see also \cite{King:1970,Keppeler:2012ih}. Second, we have to keep track of $N_\text{min}$ for each pair within the calculation as this can be higher than the intrinsic $N_\mathrm{min}$ of that pair. 
To demonstrate these points, consider the following example of multiplication by a single box, 
\begin{equation}
\label{eq:pair times box}
\ytableausetup{centerboxes,centertableaux}
\begin{aligned}\ytableausetup{centerboxes,centertableaux}
    \scalebox{0.7}{$\ydiagram[*(white)\bullet]{0,0,1+1,0+2}*[*(white)]{2+1,2+1}$} \, \otimes\, \scalebox{0.7}{$\ydiagram[*(lightgreen)]{1}$} &= 
    \scalebox{0.7}{$\ydiagram[*(white)\bullet]{0,0,1+1,0+2}*[*(white)]{2+1,2+1}*[*(lightgreen)]{3+1}$} \,\oplus\, 
    \scalebox{0.7}{$\ydiagram[*(white)\bullet]{0,0,0,1+1,0+2}*[*(lightgreen)]{0,0,2+1}*[*(white)]{2+1,2+1}$}  \,\oplus\,  
    \scalebox{0.7}{$\begin{ytableau}
    \none&\none & \empty\\
    \none &\none & \empty\\
    \none & \greendelbox{\,}\\
    \bullet &\bullet
    \end{ytableau}$}
      \,\oplus\,  
      \scalebox{0.7}{$\begin{ytableau}
    \none&\none & \empty\\
    \none &\none & \empty\\
    \none & \bullet\\
    \greendelbox{\,} &\bullet
    \end{ytableau}$}
    \\[\jot]
      &\overset{N=5}{=} 
      \scalebox{0.7}{\begin{ytableau}
      \none\\
          \none[{\begin{rotate}{45}$\scriptstyle N\sm 1$\end{rotate}}] & \none[{\begin{rotate}{45}$\scriptstyle N\sm 2$\end{rotate}}] & \none[\scriptstyle 2] & \none[\scriptstyle 1]\\
          *(lightgray)\empty & *(lightgray)\empty & \empty & *(lightgreen)\empty\\
          *(lightgray)\empty & *(lightgray)\empty & \empty\\
          *(lightgray)\empty & *(lightgray)\empty\\
          *(lightgray)\empty
      \end{ytableau}}
       \,\oplus\,  
       \scalebox{0.7}{$\begin{ytableau}
      \none\\
          \none[\text{\begin{rotate}{45}$\scriptstyle N\sm 1$\end{rotate}}] & \none[\text{\begin{rotate}{45}$\scriptstyle N\sm 2$\end{rotate}}] & \none[\scriptstyle 3] \\
          *(lightgray)\empty & *(lightgray)\empty & \empty \\
          *(lightgray)\empty & *(lightgray)\empty & \empty\\
          *(lightgray)\empty & *(lightgray)\empty & *(lightgreen)\empty\\
          *(lightgray)\empty
      \end{ytableau}$}
       \,\oplus\,  
       \scalebox{0.7}{$\begin{ytableau}
      \none\\
          \none[\text{\begin{rotate}{45}$\scriptstyle N\sm 1$\end{rotate}}] & \none[\text{\begin{rotate}{45}$\scriptstyle N\sm 1$\end{rotate}}] & \none[\scriptstyle 2] \\
          *(lightgray)\empty & *(lightgray)\empty & \empty \\
          *(lightgray)\empty & *(lightgray)\empty & \empty\\
          *(lightgray)\empty & *(lightgray)\empty \\
          *(lightgray)\empty & *(lightgreen)\empty
      \end{ytableau}$}
       \,\oplus\, 
       \scalebox{0.7}{$\begin{ytableau}
      \none\\
          \none[\scriptstyle N] & \none[\text{\begin{rotate}{45}$\scriptstyle N\sm 2$\end{rotate}}] & \none[\scriptstyle 2] \\
          \cancelbox{lightgray} & *(lightgray)\empty & \empty \\
          \cancelbox{lightgray} & *(lightgray)\empty & \empty\\
          \cancelbox{lightgray} & *(lightgray)\empty \\
          \cancelbox{lightgray} \\
          \cancelbox{lightgreen}
      \end{ytableau}$}\;.
  \end{aligned}
\end{equation}
Here the top line shows the placement of the green box in all possible positions, in the last two terms removing a barred box. We display boxes marked for removal with a dashed border, and remove them in a later step. The bottom line shows the corresponding diagrams for $N=5$, {and above each column we indicate its length for arbitrary but fixed $N$}. The gray boxes are the boxes left over when removing \raisebox{0.5ex}{\scalebox{0.6}{$\ydiagram[*(white)\bullet]{1+1,2}$}} from the corresponding $5\times2$ rectangle. Since the boxes in \raisebox{0.5ex}{\scalebox{0.6}{$\ydiagram[*(white)\bullet]{1+1,2}$}} label the boxes \textit{not} in the fixed-$N$ version of the pair, 
appending a box to a gray column is equivalent to \textit{removing} a box from \raisebox{1ex}{\scalebox{0.6}{$\ydiagram[*(white)\bullet]{1+1,2}$}}.

Next, consider the third term in \Eqref{eq:pair times box}, and note that the dashed box in the upper right corner of the barred diagram corresponds to an \textit{existing} box in the corresponding fixed-$N$ diagram. Therefore, when multiplying a diagram pair by a more complicated Young diagram (rather than a single box) it is possible to add boxes from subsequent columns to the right of this dashed box without increasing the number of rows of the pair.

To see how the $N$-dependence appears, observe the second last diagram pair in the first line of \Eqref{eq:pair times box} and its fixed-$N$ version in the second line. When we remove the {dashed} box, we retain the diagram pair
\raisebox{0.5ex}{\scalebox{0.6}{\ytableausetup{centerboxes,centertableaux}\begin{ytableau}
    \none&\none & \empty\\ \none &\none & \empty\\ \bullet &\bullet
\end{ytableau}}} with intrinsic $N_\mathrm{min}=3$. Correspondingly, the fixed-$N$ diagram exists for all $N\geq3$. However, within the equation, $N_\mathrm{min}$ for this term is $4$, since before removing the box, the diagram pair already had $N_\mathrm{min}=4$. In fact, in this example the whole equation has $N_\mathrm{min}=4$, since the first factor on the left-hand side has $N_\mathrm{min}=4$.

When multiplying pairs by diagrams with more than one box it can even happen that the $N_\mathrm{min}$ value of a term is not only higher than the intrinsic $N_\mathrm{min}$ of that diagram pair but also higher than the $N_\mathrm{min}$ values of both factors on the left-hand side of the equation. Consider
\begin{equation}\label{eq:pair_times_11}
\ytableausetup{centerboxes,centertableaux}
\scalebox{0.7}{$\begin{ytableau}
            \none & \empty \\
            \bullet
        \end{ytableau}$}
\,\otimes\,
\scalebox{0.7}{$\begin{ytableau}
           *(lightgreen)\empty\\
            *(lightgreen)\empty
        \end{ytableau}$}
=
\scalebox{0.7}{$\begin{ytableau}
            \none & \empty & *(lightgreen)\empty\\
           \none & *(lightgreen)\empty\\
            \bullet
        \end{ytableau}$}
\,\oplus\,
\scalebox{0.7}{$\begin{ytableau}
            \none & \empty & *(lightgreen)\empty\\
            \greendelbox{\,}
        \end{ytableau}$}
\,\oplus\,
\scalebox{0.7}{$\begin{ytableau}
            \none & \empty \\
            \none & *(lightgreen)\empty\\
            \none & *(lightgreen)\empty\\
            \bullet
        \end{ytableau}$}
\,\oplus\,
\scalebox{0.7}{$\begin{ytableau}
            \none & \empty \\
            \none & *(lightgreen)\empty\\
            \greendelbox{\,}
        \end{ytableau}$}\;,
\end{equation}
where (for example) the last term has an intrinsic $N_\mathrm{min}$ of $2$ but it here appears with $N_\mathrm{min}=3$, since the row needed for the soon to-be-removed {dashed} box has to be taken into account. 
This term is also absent in the $N=2$ version of this equation,
\begin{equation}
\ytableausetup{centerboxes,centertableaux}
\scalebox{0.7}{$\begin{ytableau}
            \none & \empty \\
            \bullet
        \end{ytableau}$}
\,\otimes\,
\scalebox{0.7}{$\ydiagram{1,1}$}
\,\overset{N=2}{=}\,
\scalebox{0.7}{$\ydiagram{2}$}
\,\otimes\,
\bullet
\,=\,
\scalebox{0.7}{$\ydiagram{2}$}\;.
\end{equation}
Here and in the following
$\bullet$ denotes the empty Young diagram, i.e., the trivial
representation (not to be confused with boxes \raisebox{.5ex}{{\tiny$\begin{ytableau} \bullet \end{ytableau}$}} of the barred diagram).  

Hence, we have to keep track of $N_\mathrm{min}$ for each term individually, and we do so by writing it as a subscript to the multiplicities in front of Young diagrams or diagram pairs. With this notation, and after removing {dashed} boxes, \Eqref{eq:pair_times_11} reads
\begin{equation}
\scalebox{0.7}{$\begin{ytableau}
            \none & \empty \\
            \bullet
        \end{ytableau}$}
\,\otimes\,
\scalebox{0.7}{$\begin{ytableau}
           \empty\\
           {\empty}
        \end{ytableau}$}
\,=\,
1_3\,
\scalebox{0.7}{$\begin{ytableau}
            \none & \empty & \empty\\
           \none & \empty\\
            \bullet
        \end{ytableau}$}
\,\oplus\,
1_2\,
\scalebox{0.7}{$\begin{ytableau}
            \empty & \empty
        \end{ytableau}$}
\,\oplus\,
1_4\,
\scalebox{0.7}{$\begin{ytableau}
            \none & \empty \\
            \none & \empty\\
            \none & \empty\\
            \bullet
        \end{ytableau}$}
\,\oplus\,
1_3\,
\scalebox{0.7}{$\begin{ytableau}
            \empty \\
            {\empty}
        \end{ytableau}$}\;.
\end{equation}
After these observations, we are ready to state the general algorithm for column multiplication of a Young diagram pair and an ordinary Young diagram:
\begin{theorem}[Multiplying a Young diagram pair by an ordinary Young diagram]
\label{alg:pair_diagram_mult}
    The Young diagram pairs resulting from the product $\pair{\rho}{\sigma}\otimes {\mu}$ are obtained via the following procedure:
\begin{enumerate}
    \item Label each box in $\mu$ by the column in which it appears.
    \item Starting from the first, i.e., left-most, column in ${\mu}$, add all boxes from column $k$, each labeled ${k}$, to a semitableau pair $\pair{\gamma}{\delta}$ from a previous step, so that $\delta$ is extended (with solid-border boxes), or boxes from $\overline{\gamma}$ are removed (indicated by dashed borders). 
    In this step, treat dashed boxes from previous steps as if they had been removed. Keep the resulting semitableau pair if it
        \begin{enumerate}
            \item corresponds to a Young diagram pair, i.e., when dashed boxes are removed, the result has the shape of a Young diagram pair,
            \item has no two boxes labeled ${k}$ appearing in the same {row}, 
            \item has a column-sequence of numbers, read bottom-to-top and then left-to-right, that is admissible.
        \end{enumerate}
    \item After exhausting the columns of $\mu$, for each semitableau pair, determine $N_\text{min}$ as the number of rows in that semitableau pair.
    \item Remove dashed boxes and all labels, and collect the $N$-dependent multiplicities for each diagram pair.
    
\end{enumerate}
\end{theorem}

We illustrate this algorithm with another example, in which the
ordinary diagram has more than one column, and demonstrate under which circumstances dashed and solid boxes can end up in the same row.
Consider 
\begin{align}\ytableausetup{centerboxes,centertableaux}\label{eq:pair_diagram_mult_example}
  \scalebox{0.7}{$\ydiagram[*(white)\bullet]{0,1}*[*(white)]{1+1}$}
  \otimes 
  \scalebox{0.7}{$\ydiagram{2,1} $}
  =& \ \ytableausetup{centerboxes,centertableaux}
  \scalebox{0.7}{$\ydiagram[*(white)\bullet]{0,1}*[*(white)]{1+1}$}
  \otimes
  \scalebox{0.7}{$\begin{ytableau}
      1 & 2\\
      1
    \end{ytableau}$}\\
  =& \ytableausetup{centerboxes,centertableaux}\left(
  \scalebox{0.7}{$\begin{ytableau}
      \none & \empty & 1\\
      \none & 1 \\
      \bullet 
    \end{ytableau}$}
  \,\oplus\,
  \scalebox{0.7}{$\begin{ytableau}
      \none & \empty & 1\\
      \delbox{1} 
    \end{ytableau}$}
    \,\oplus\,
  \scalebox{0.7}{$\begin{ytableau}
      \none & \empty\\
      \none & 1 \\
      \none & 1 \\
      \bullet 
    \end{ytableau}$}
  \,\oplus\,
  \scalebox{0.7}{$\begin{ytableau}
      \none & \empty\\
      \none & 1 \\
      \delbox{1} 
    \end{ytableau}$}
  \right)\otimes\scalebox{0.7}{$\ytableaushort{2}$}\nonumber\\
  =& \ytableausetup{centerboxes,centertableaux}\ 
  \scalebox{0.7}{$\begin{ytableau}
      \none & \empty & 1 & 2\\
      \none & 1 \\
      \bullet 
    \end{ytableau}$}
  \,\oplus\,
  \scalebox{0.7}{$\begin{ytableau}
      \none & \empty & 1\\
      \none & 1  & 2\\
      \bullet 
    \end{ytableau}$}
  \,\oplus\,
  \scalebox{0.7}{$\begin{ytableau}
      \none & \empty & 1 & 2\\
      \delbox{1} 
    \end{ytableau}$}
  \,\oplus\,
  \scalebox{0.7}{$\begin{ytableau}
      \none & \empty & 1\\
      \delbox{1} & 2  
    \end{ytableau}$}
    \,\oplus\,
  \scalebox{0.7}{$\begin{ytableau}
      \none & \empty & 2\\
      \none & 1 \\
      \none & 1 \\
      \bullet 
    \end{ytableau}$}
  \,\oplus\,
  \scalebox{0.7}{$\begin{ytableau}
      \none & \empty & 2\\
      \none & 1 \\
      \delbox{1}  
    \end{ytableau}$}
  \,\oplus\,
  \scalebox{0.7}{$\begin{ytableau}
      \none & \empty\\
      \none & 1 \\
      \delbox{1}  & 2  
    \end{ytableau}$}\ ,
  \nonumber
\end{align}
where we have already rejected terms like \raisebox{0.5ex}{\scalebox{0.6}{\ytableausetup{centerboxes,centertableaux}\begin{ytableau} \none & \empty & 1\\ \none& 1\\\delbox{2} \end{ytableau}}} due to non-admissibility of the column sequence, but we have still kept dashed boxes.  Note that ordinary boxes can \textit{only} appear in the same row as removed barred boxes (as in the fourth and in the last term) if the barred box has been removed in a step \textit{before} the ordinary box is added, e.g., the pattern
\raisebox{0.5ex}{\scalebox{0.6}{\ytableausetup{centerboxes,centertableaux}\begin{ytableau} \delbox{1} & 2 \end{ytableau}}}
can occur, but not
\raisebox{0.5ex}{\scalebox{0.6}{\ytableausetup{centerboxes,centertableaux}\begin{ytableau} \delbox{2} & 1 \end{ytableau}}}. In the final step we write down $N_\mathrm{min}$ for each semitableau and then remove dashed boxes (as well as all box labels),
\begin{align}
  \label{eq:pair_diagram_mult_example_2}
  \scalebox{0.7}{$\ydiagram[*(white)\bullet]{0,1}*[*(white)]{1+1}$}
  \otimes 
  \scalebox{0.7}{$\ydiagram{2,1}$}
  \,=\,
  \ytableausetup{centerboxes,centertableaux}\ 
  1_3
  \scalebox{0.7}{$\begin{ytableau}
      \none & \empty & \empty & \empty\\
      \none & \empty \\
      \bullet 
    \end{ytableau}$}
  \oplus
  1_3
  \scalebox{0.7}{$\begin{ytableau}
      \none & \empty & \empty\\
      \none & \empty  & \empty\\
      \bullet 
    \end{ytableau}$}
  \oplus
  1_2\, \scalebox{0.7}{$\ydiagram{3}$}
  \oplus (1_2+1_3)\, \scalebox{0.7}{$\ydiagram{2,1}$}
  \oplus
  1_4
  \scalebox{0.7}{$\begin{ytableau}
      \none & \empty & \empty\\
      \none & \empty \\
      \none & \empty \\
      \bullet 
    \end{ytableau}$} 
  \oplus 1_3\, \scalebox{0.7}{$\ydiagram{1,1,1}$}
  \, .
\end{align}
Note that the fourth and the sixth term in \Eqref{eq:pair_diagram_mult_example} yield the same final diagram but with different $N_\mathrm{min}$ values, i.e., for $N=2$ the diagram $\raisebox{0.5ex}{\scalebox{0.6}{$\ydiagram{2,1}$}}\overset{N=2}{=}\raisebox{0.5ex}{\scalebox{0.6}{$\ydiagram{1}$}}$ appears with multiplicity one in this decomposition, whereas for $N\geq3$ it appears with multiplicity two.

We conclude this subsection by remarking that for $N=3$, \Eqref{eq:pair_diagram_mult_example_2} becomes a version of the well-known $8\otimes 8$ decomposition,
\begin{equation}
  \label{eq:pair_diagram_mult_example_N3}
  \scalebox{0.7}{$\ydiagram[*(white)\bullet]{0,1}*[*(white)]{1+1}$}
  \otimes 
  \scalebox{0.7}{$\ydiagram{2,1}$}
  \,\overset{N=3}{=}\,
  \underbrace{
    \scalebox{0.7}{$\ydiagram{2,1}$}}_{8}
  \otimes \underbrace{
    \scalebox{0.7}{$\ydiagram{2,1}$}}_{8}
  \,=\, 
  \underbrace{
    \scalebox{0.7}{$\ydiagram{4,2}$}}_{27}\,
  \oplus\,
  \underbrace{
    \scalebox{0.7}{$\ydiagram{3,3}$}}_{\overline{10}}\,
  \oplus\,
  \underbrace{\scalebox{0.7}{$\ydiagram{3}$}}_{10}\,
  \oplus\,2\ 
  \underbrace{
    \scalebox{0.7}{$\ydiagram{2,1}$}}_{8}\,
  \oplus\,
  \underbrace{\bullet }_{1}\
  \;,    
\end{equation}
where $\bullet$ again denotes the empty diagram, i.e., the trivial representation.

\subsection{Multiplying a Young diagram pair by a barred Young diagram}
\label{sec:barred multiplication}

Having discussed multiplication of diagram pairs by ordinary Young diagrams, we now study multiplication of diagram pairs by barred Young diagrams. 
We could straightforwardly obtain an algorithm for this by conjugating the above algorithm, as $\pair{\rho}{\sigma}\otimes \overline{\mu}$ is equivalent to $\overline{\pair{\sigma}{\rho}\otimes \mu}$. 
However, as our goal is to derive an algorithm for pair-pair multiplication, we separately derive an algorithm involving only a barred diagram, in order to later combine it with  Algorithm~\ref{alg:pair_diagram_mult}.
To this end, we again map diagram pairs, and now also barred diagrams, to ordinary Young diagrams with $N$-dependent column lengths, then multiply those using column multiplication (Algorithm~\ref{alg:column_multiplication}), and finally map the result back to $N$-independent diagram pairs. 

First we consider multiplication of a diagram pair by a single barred box, 
\begin{equation}\label{eq:Nc3_adding_box}
    \begin{aligned}\ytableausetup{centerboxes,centertableaux}
         \scalebox{0.7}{$\ydiagram{1+2,1+1}*[\bullet]{0,0,1,1}$}\,\otimes\,
         \scalebox{0.7}{$\ydiagram[\bullet]{1}$} \ \overset{N=5}{=}& \ 
         \scalebox{0.7}{$\begin{ytableau}
          \none[\text{\begin{rotate}{45}$\scriptstyle N\sm 2$\end{rotate}}] & \none[\scriptstyle 2] & \none[\scriptstyle 1]\\
          *(lightgray)\empty & \empty & \empty \\
          *(lightgray)\empty & \empty \\
          *(lightgray)\empty 
      \end{ytableau}$}
      \,\otimes\,
      \scalebox{0.7}{$\begin{ytableau}
          \none[\text{\begin{rotate}{45}$\scriptstyle N\sm 1$\end{rotate}}] \\
          *(lightgreen)\empty \\
          *(lightgreen)\empty \\
          *(lightgreen)\empty \\
          *(lightgreen)\empty
      \end{ytableau} $}
      \ = \ 
         \scalebox{0.7}{$\begin{ytableau}
          \none[\text{\begin{rotate}{45}$\scriptstyle N\sm 1$\end{rotate}}] & \none[\text{\begin{rotate}{45}$\scriptstyle N\sm 2$\end{rotate}}] & \none[\scriptstyle 2] & \none[\scriptstyle 1]\\
          *(lightgray)\empty & \empty & \empty &*(lightgreen)\empty \\
          *(lightgray)\empty & \empty &*(lightgreen)\empty\\
          *(lightgray)\empty &*(lightgreen)\empty\\
          *(lightgreen)\empty
      \end{ytableau}$}
      \,\oplus\,
      \scalebox{0.7}{$\begin{ytableau}
          \none[\scriptstyle N] & \none[\text{\begin{rotate}{45}$\scriptstyle N\sm 3$\end{rotate}}] & \none[\scriptstyle 2] & \none[\scriptstyle1]\\
          *(lightgray) & \empty & \empty & *(lightgreen) \empty\\
          *(lightgray) & \empty &*(lightgreen)\empty\\
          *(lightgray) \\
          *(lightgreen)\\
          *(lightgreen)
      \end{ytableau}$}
       \,\oplus \,
      \scalebox{0.7}{$\begin{ytableau}
          \none[\scriptstyle N] & \none[\text{\begin{rotate}{45}$\scriptstyle N\sm 2$\end{rotate}}] & \none[\scriptstyle 1] & \none[\scriptstyle 1]\\
          *(lightgray) & \empty & \empty  &*(lightgreen)\empty\\
          *(lightgray) & \empty \\
          *(lightgray) &*(lightgreen)\empty\\
          *(lightgreen)\\
          *(lightgreen)
      \end{ytableau}$}
      \,\oplus\,
         \scalebox{0.7}{$\begin{ytableau}
          \none[\scriptstyle N] & \none[\text{\begin{rotate}{45}$\scriptstyle N\sm 2$\end{rotate}}] & \none[\scriptstyle 2] \\
          *(lightgray) & \empty & \empty  \\
          *(lightgray) & \empty &*(lightgreen)\empty\\
          *(lightgray) &*(lightgreen)\empty\\
          *(lightgreen)\\
          *(lightgreen)
      \end{ytableau}$}\;,
    \end{aligned}
\end{equation}
where, as in \Eqref{eq:pair times box}, we display $N$-dependent diagrams for a particular $N$, here $N=5$, but also keep track of the length of each column for general $N$, which will allow us to later translate back to $N$-independent diagram pairs.
We remark that the second column in the first term on the right-hand side, which here has 3 boxes, generally comes with $N{-}2$ boxes, and correspondingly for the second columns of the other diagrams. The reason is that for a higher $N$, many, more precisely $N{-}2{-}2$, boxes are added to that column. More generally we note that, since the two $N$-dependent diagrams which we multiply consist of in total $2N$ boxes, each term in the direct sum also consists of $2N$ boxes, i.e., it must have $2$ columns with $N$-dependent column lengths. By the same argument, the number of $N$-dependent columns before and after applying Algorithm~\ref{alg:column_multiplication} (column multiplication) is always conserved as long as we refrain from discarding columns of length $N$.

Term-by-term mapping back to diagram pairs, we observe that in some terms multiplication by a barred box has the effect of removing a box from the ordinary part, whereas in other terms a box is added to the barred part, cf.\ \cite{Sjodahl:2018cca},
\begin{equation}\label{eq:Nc_general_adding_box}
\ytableausetup{mathmode,centerboxes, centertableaux}
\begin{split}
    \scalebox{0.7}{$\ydiagram{1+2,1+1}*[\bullet]{0,0,1,1}$}
    \,\otimes\,
    \scalebox{0.7}{$\ydiagram[*(lightgreen)\bullet]{1}$} &= 
    \scalebox{0.7}{$\ydiagram{2+2,2+1}*[\bullet]{0,0,1+1,1+1}*[*(lightgreen)\bullet]{0,0,0,1}$}
    \,\oplus\, 
    \scalebox{0.7}{$ \ydiagram{1+2,1+1}*[\bullet]{0,0,0,1,1}*[*(lightgreen)\bullet]{0,0,1}$}
    \,\oplus\, 
    \scalebox{0.7}{$\begin{ytableau}
        \none & \empty & \empty\\
        \none & \greendelbox{\,}\\
        \bullet\\\bullet
    \end{ytableau}$}
    \,\oplus \,
    \scalebox{0.7}{$\begin{ytableau}
        \none & \empty & \greendelbox{\,}\\
        \none & \empty\\
        \bullet\\\bullet
    \end{ytableau}$}
    \\
    &= 1_4\,\scalebox{0.7}{$\ydiagram{2+2,2+1}*[\bullet]{0,0,1+1,1+1}*[\bullet]{0,0,0,1}$}
     \,\oplus\, 
     1_5\,\scalebox{0.7}{$\ydiagram{1+2,1+1}*[\bullet]{0,0,0,1,1}*[\bullet]{0,0,1}$}
    \,\oplus\, 
     1_4\,\scalebox{0.7}{$\ydiagram{1+2}*[\bullet]{0,1,1}$} 
    \,\oplus \,
    1_4\,\scalebox{0.7}{$\ydiagram{1+1,1+1}*[\bullet]{0,0,1,1}$}\;.
    \end{split}
\end{equation}
As in \secref{sec:pair x ordinary} we first display removed boxes with dashed border, and in the second line, after writing down $N_\mathrm{min}$ for each diagram pair (the number of rows of that pair in the line above), we remove the dashed boxes.

According to Algorithm~\ref{alg:column_multiplication} (column multiplication) in each term  in \Eqref{eq:Nc3_adding_box} we can add at most one green box to every row. The row to which we add no green box in \Eqref{eq:Nc3_adding_box} corresponds to the row to which we either add  a green barred box \raisebox{0.5ex}{\scalebox{0.6}{$\ydiagram[*(lightgreen)\bullet]{1}$}} or from which we remove an ordinary box \raisebox{0.75ex}{\scalebox{0.6}{$\begin{ytableau}\greendelbox{\,}\end{ytableau}$}} in \Eqref{eq:Nc_general_adding_box}. 
By considering the result in pair notation, we observe that barred boxes are placed above and/or to the left of the boxes already present in the pair (or they remove some ordinary boxes), in contrast to ordinary boxes that are placed below and/or to the right of those already in the pair (or they remove some barred boxes).

To illustrate what happens when multiplying by a barred diagram with longer columns and with several columns, consider
\vspace{-5pt}
\begin{equation}\label{eq:pair x 21bar N-dep}
    \begin{split}
        \ytableausetup{mathmode, centerboxes,centertableaux, boxframe=normal}
            &\scalebox{0.7}{$\ydiagram[*(white)\bullet]{0,1}*{1+1}$}
            \,\otimes \,
            \scalebox{0.7}{$\ydiagram[*(white)\bullet]{1+1,2}$}
            \ \overset{N=4}{=} \ 
            \ytableausetup{mathmode, centerboxes,centertableaux, boxframe=normal}\ 
            \scalebox{0.7}{$\begin{ytableau} 
    \none[\text{\begin{rotate}{45}$\scriptstyle N\sm 1$\end{rotate}}] & \none[\scriptstyle 1]\\
            \empty & \empty \\
            \empty \\
            \empty  \\
            \end{ytableau}$}
            \,\otimes\, 
            \scalebox{0.7}{$\begin{ytableau} 
    \none[\text{\begin{rotate}{45}$\scriptstyle N\sm 1$\end{rotate}}] & \none[\text{\begin{rotate}{45}$\scriptstyle N\sm 2$\end{rotate}}]\\
            *(lightgreen)\sm 2 & *(lightblue) \sm 1\\
            *(lightgreen)\sm 2 & *(lightblue) \sm 1\\
            *(lightgreen)\sm 2 \\
            \end{ytableau}$}
            \ = \ \left(
            \scalebox{0.7}{$\begin{ytableau} 
    \none[\text{\begin{rotate}{45}$\scriptstyle N\sm 1$\end{rotate}}] & \none[\text{\begin{rotate}{45}$\scriptstyle N\sm 1$\end{rotate}}] & \none[\text{$\scriptstyle 1$}]\\
            \empty & \empty & *(lightgreen)\sm 2\\
            \empty & *(lightgreen)\sm 2\\
            \empty & *(lightgreen)\sm 2 \\
            \end{ytableau}$}
            \,\oplus\,
            \scalebox{0.7}{$\begin{ytableau} 
    \none[\text{$\scriptstyle N$}] & \none[\text{\begin{rotate}{45}$\scriptstyle N\sm 2$\end{rotate}}] & \none[\text{$\scriptstyle 1$}]\\
            \empty & \empty & *(lightgreen)\sm 2\\
            \empty & *(lightgreen)\sm 2\\
            \empty \\
            *(lightgreen)\sm 2 \\
            \end{ytableau}$}
            \,\oplus\,
            \scalebox{0.7}{$\begin{ytableau} 
    \none[\text{$\scriptstyle N$}] & \none[\text{\begin{rotate}{45}$\scriptstyle N\sm 1$\end{rotate}}] \\
            \empty & \empty \\
            \empty & *(lightgreen)\sm 2\\
            \empty & *(lightgreen)\sm 2\\
            *(lightgreen)\sm 2 \\
            \end{ytableau}$}
            \;\right)
            \otimes
             \scalebox{0.7}{$\begin{ytableau} 
    \none[\text{\begin{rotate}{45}$\scriptstyle N\sm 2$\end{rotate}}]\\
            *(lightblue) \sm 1\\
            *(lightblue) \sm 1\\
               \end{ytableau}$}
            \\[2ex]
            &\ = \  
            \ytableausetup{mathmode, centerboxes,centertableaux, boxframe=normal}
            \scalebox{0.7}{$\begin{ytableau} 
    \none[\text{\begin{rotate}{45}$\scriptstyle N\sm 1$\end{rotate}}] & \none[\text{\begin{rotate}{45}$\scriptstyle N\sm 1$\end{rotate}}] &
    \none[\text{\begin{rotate}{45}$\scriptstyle N\sm 2$\end{rotate}}] & \none[\text{$\scriptstyle 1$}]\\
            \empty & \empty & *(lightgreen)\sm 2 & *(lightblue) \sm 1\\
            \empty & *(lightgreen)\sm 2 & *(lightblue) \sm 1\\
            \empty & *(lightgreen)\sm 2 \\
            \end{ytableau}$}
            \,\oplus\,
            \scalebox{0.7}{$\begin{ytableau} 
    \none[\text{\begin{rotate}{45}$\scriptstyle N\sm 1$\end{rotate}}] & 
    \none[\text{\begin{rotate}{45}$\scriptstyle N\sm 1$\end{rotate}}] & \none[\text{\begin{rotate}{45}$\scriptstyle N\sm 1$\end{rotate}}] \\
            \empty & \empty & *(lightgreen)\sm 2 \\
            \empty & *(lightgreen)\sm 2 & *(lightblue) \sm 1\\
            \empty & *(lightgreen)\sm 2 & *(lightblue) \sm 1\\
            \end{ytableau}$}
            \,\oplus\,
            \scalebox{0.7}{$\begin{ytableau} 
    \none[\text{$\scriptstyle N$}] & 
    \none[\text{\begin{rotate}{45}$\scriptstyle N\sm 2$\end{rotate}}] & \none[\text{\begin{rotate}{45}$\scriptstyle N\sm 2$\end{rotate}}] & \none[\text{$\scriptstyle 1$}]\\
            \empty & \empty & *(lightgreen)\sm 2 & *(lightblue) \sm 1\\
            \empty & *(lightgreen)\sm 2 & *(lightblue) \sm 1\\
            \empty \\
            *(lightgreen)\sm 2 \\
            \end{ytableau}$}
            \,\oplus\,
            \scalebox{0.7}{$\begin{ytableau} 
    \none[\text{$\scriptstyle N$}] & 
    \none[\text{\begin{rotate}{45}$\scriptstyle N\sm 1$\end{rotate}}] & \none[\text{\begin{rotate}{45}$\scriptstyle N\sm 3$\end{rotate}}] & \none[\text{$\scriptstyle 1$}]\\
            \empty & \empty & *(lightgreen)\sm 2 & *(lightblue) \sm 1\\
            \empty & *(lightgreen)\sm 2\\
            \empty  & *(lightblue) \sm 1\\
            *(lightgreen)\sm 2 \\
            \end{ytableau}$}
            \,\oplus\,
            \scalebox{0.7}{$\begin{ytableau} 
    \none[\text{$\scriptstyle N$}] & 
    \none[\text{\begin{rotate}{45}$\scriptstyle N\sm 1$\end{rotate}}] & \none[\text{\begin{rotate}{45}$\scriptstyle N\sm 2$\end{rotate}}] \\
            \empty & \empty & *(lightgreen)\sm 2 \\
            \empty & *(lightgreen)\sm 2 & *(lightblue) \sm 1\\
            \empty  & *(lightblue) \sm 1\\
            *(lightgreen)\sm 2 \\
            \end{ytableau}$}
            \,\oplus\,
            \scalebox{0.7}{$\begin{ytableau} 
    \none[\text{$\scriptstyle N$}] & 
    \none[\text{\begin{rotate}{45}$\scriptstyle N\sm 1$\end{rotate}}] & \none[\text{\begin{rotate}{45}$\scriptstyle N\sm 2$\end{rotate}}] \\
            \empty & \empty & *(lightblue) \sm 1\\
            \empty & *(lightgreen)\sm 2 & *(lightblue) \sm 1\\
            \empty & *(lightgreen)\sm 2\\
            *(lightgreen)\sm 2 \\
            \end{ytableau}$}
            \,\oplus\,
            \scalebox{0.7}{$\begin{ytableau} 
    \none[\text{$\scriptstyle N$}] & 
    \none[\text{$\scriptstyle N$}] & 
    \none[\text{\begin{rotate}{45}$\scriptstyle N\sm 3$\end{rotate}}] \\
            \empty & \empty & *(lightblue) \sm 1\\
            \empty & *(lightgreen)\sm 2 \\
            \empty & *(lightgreen)\sm 2\\
            *(lightgreen)\sm 2 & *(lightblue) \sm 1\\
            \end{ytableau}$}\;.
    \end{split}
\end{equation}
Here we choose to label boxes of columns of the $N$-dependent image of the barred diagram by negative integers $\sm k,\ldots,\sm 1$. Using this labeling, our application of Algorithm~\ref{alg:column_multiplication} (column multiplication) does not change, i.e., we have the same admissibility condition. Later on, when generalizing to multiplication by diagram pairs, we reserve positive integers for columns coming from the ordinary part of the pair, and negative integers for columns coming from the barred part of the pair. Mapping the result back to pair notation, i.e., replacing columns of length $N{-}\ell$ by barred columns of length $\ell$, we find
\begin{equation}\label{eq:pair x 21bar color}
\ytableausetup{mathmode, centerboxes,centertableaux, boxframe=normal}
\scalebox{0.7}{$\ydiagram[*(white)\bullet]{0,1}*{1+1}$}
            \,\otimes \,
    \scalebox{0.7}{$\begin{ytableau}
\none & *(lightblue)\bullet\\
*(lightgreen) \bullet & *(lightblue)\bullet\\
\end{ytableau}$}
            =
\scalebox{0.7}{$\begin{ytableau}
\none & \none & \none & \empty\\
\none & \none & *(lightblue)\bullet \\
*(lightblue)\bullet & *(lightgreen)\bullet & \bullet & \none
\end{ytableau}$}
\,\oplus\,
\scalebox{0.7}{$\begin{ytableau}
\none & \none & \none & \colourdelbox{lightblue}{\,}\\
*(lightblue)\bullet & *(lightgreen)\bullet & \bullet & \none
\end{ytableau}$}
\,\oplus\,
\scalebox{0.7}{$\begin{ytableau}
\none & \none & \empty \\
*(lightblue)\bullet & *(lightgreen)\bullet\\
*(lightblue)\bullet & \bullet 
\end{ytableau}$}
\,\oplus\,
\scalebox{0.7}{$\begin{ytableau}
\none & \none & \\
\none &*(lightblue)\bullet\\
\none &*(lightgreen)\bullet\\
*(lightblue)\bullet &\bullet
\end{ytableau}$}
\,\oplus\,
\scalebox{0.7}{$\begin{ytableau}
\none & \none & \colourdelbox{lightblue}{\,}\\
\none & *(lightgreen)\bullet\\
*(lightblue) \bullet &\bullet
\end{ytableau}$}
\,\oplus\,
\scalebox{0.7}{$\begin{ytableau}
\none & *(lightblue)\bullet & \colourdelbox{lightgreen}{\,}\\
*(lightblue)\bullet &\bullet
\end{ytableau}$}
\,\oplus\,
\scalebox{0.7}{$\begin{ytableau}
*(lightblue)\bullet & \colourdelbox{lightgreen}{\,}\\
*(lightblue)\bullet\\
\bullet
\end{ytableau}$}\,,
\end{equation}
where we again display removed boxes with dashed borders. In order to explain the coloring we also map the intermediate result, after just adding boxes labeled $\sm2$, back to pair notation,
\begin{equation}\label{eq:map back intermediate}
\ytableausetup{mathmode, centerboxes,centertableaux, boxframe=normal}
\scalebox{0.7}{$\begin{ytableau} 
    \none[\text{\begin{rotate}{45}$\scriptstyle N\sm 1$\end{rotate}}] & \none[\text{\begin{rotate}{45}$\scriptstyle N\sm 1$\end{rotate}}] & \none[\text{$\scriptstyle 1$}]\\
    \empty & \empty & *(lightgreen)\sm 2\\
    \empty & *(lightgreen)\sm 2\\
    \empty & *(lightgreen)\sm 2 \\
  \end{ytableau}$}
\,\oplus\,
\scalebox{0.7}{$\begin{ytableau} 
    \none[\text{$\scriptstyle N$}] & \none[\text{\begin{rotate}{45}$\scriptstyle N\sm 2$\end{rotate}}] & \none[\text{$\scriptstyle 1$}]\\
    \empty & \empty & *(lightgreen)\sm 2\\
    \empty & *(lightgreen)\sm 2\\
    \empty \\
    *(lightgreen)\sm 2 \\
  \end{ytableau}$}
\,\oplus\,
\scalebox{0.7}{$\begin{ytableau} 
    \none[\text{$\scriptstyle N$}] & \none[\text{\begin{rotate}{45}$\scriptstyle N\sm 1$\end{rotate}}] \\
    \empty & \empty \\
    \empty & *(lightgreen)\sm 2\\
    \empty & *(lightgreen)\sm 2\\
    *(lightgreen)\sm 2 \\
  \end{ytableau}$}
\quad \mapsto \quad 
\scalebox{0.7}{$\begin{ytableau}
    \none & \none & \empty\\
    *(lightgreen)\bullet & \bullet & \none
  \end{ytableau}$}
\,\oplus\,
\scalebox{0.7}{$\begin{ytableau}
    \none & \empty \\
    *(lightgreen)\bullet\\
    \bullet 
  \end{ytableau}$}
\,\oplus\,
\scalebox{0.7}{$\begin{ytableau}
    \none & \colourdelbox{lightgreen}{\,}\\
    \bullet
  \end{ytableau}$}\,.
\end{equation}
This also explains the different placement of the dashed boxes in \Eqref{eq:pair x 21bar color}, in terms 2 and 5 in its own row, and in terms 6 and 7 sharing a row with a box from the barred part of the diagram, since in the last two terms the box is deleted \textit{before} the barred diagram is extended by blue barred boxes \raisebox{.5ex}{\scalebox{0.6}{$\ydiagram[*(lightblue)\bullet]{1}$}}.

Consequently, the fifth and sixth terms in \eqref{eq:pair x 21bar color}, which both become \raisebox{0.5ex}{\scalebox{0.6}{$\ydiagram[*(white)\bullet]{1+1,2}$}} after removing the dashed box, have $N_\mathrm{min}=3$ and $N_\mathrm{min}=2$, respectively, since the fifth term derives from the middle term in \Eqref{eq:map back intermediate} which does not exist for $N<3$, whereas the sixth term derives from the last term in \Eqref{eq:map back intermediate} which also exists for $N=2$. 

For a systematic treatment of multiplication by arbitrary barred diagrams we label boxes of the barred diagram by barred numbers $\overline{1},\overline{2},\ldots$ according to the column in which the box appears, counting columns right-to-left. Example~\eqref{eq:pair x 21bar color} then reads
\begin{equation}\label{eq:pair x 21bar color barred labels}
\ytableausetup{mathmode, centerboxes,centertableaux, boxframe=normal}
\scalebox{0.7}{$\ydiagram[*(white)\bullet]{0,1}*{1+1}$}
            \,\otimes \,
    \scalebox{0.7}{$\begin{ytableau}
\none & *(lightblue)\overline{1}\\
*(lightgreen) \overline{2} & *(lightblue)\overline{1}\\
\end{ytableau}$}
            =
\scalebox{0.7}{$\begin{ytableau}
\none & \none & \none & \empty\\
\none & \none & *(lightblue)\overline{1} \\
*(lightblue)\overline{1} & *(lightgreen) \overline{2} & \bullet & \none
\end{ytableau}$}
\,\oplus\,
\scalebox{0.7}{$\begin{ytableau}
\none & \none & \none & \colourdelbox{lightblue}{\overline{1}}\\
*(lightblue)\overline{1} & *(lightgreen) \overline{2} & \bullet & \none
\end{ytableau}$}
\,\oplus\,
\scalebox{0.7}{$\begin{ytableau}
\none & \none & \empty \\
*(lightblue)\overline{1} & *(lightgreen) \overline{2}\\
*(lightblue)\overline{1} & \bullet 
\end{ytableau}$}
\,\oplus\,
\scalebox{0.7}{$\begin{ytableau}
\none & \none & \\
\none &*(lightblue)\overline{1}\\
\none &*(lightgreen) \overline{2}\\
*(lightblue)\overline{1} &\bullet
\end{ytableau}$}
\,\oplus\,
\scalebox{0.7}{$\begin{ytableau}
\none & \none & \colourdelbox{lightblue}{\overline{1}}\\
\none & *(lightgreen) \overline{2}\\
*(lightblue) \overline{1} &\bullet
\end{ytableau}$}
\,\oplus\,
\scalebox{0.7}{$\begin{ytableau}
\none & *(lightblue)\overline{1} & \colourdelbox{lightgreen}{\overline{2}}\\
*(lightblue)\overline{1} &\bullet
\end{ytableau}$}
\,\oplus\,
\scalebox{0.7}{$\begin{ytableau}
*(lightblue)\overline{1} & \colourdelbox{lightgreen}{\overline{2}}\\
*(lightblue)\overline{1}\\
\bullet
\end{ytableau}$}
\end{equation}
where we display the now obsolete colors for a last time. Note that boxes with higher barred labels \raisebox{.5ex}{\scalebox{0.7}{$\ytableausetup{mathmode, centerboxes,centertableaux, boxframe=normal}\begin{ytableau}*(lightgreen) \overline{2}\end{ytableau}$}} are added before boxes with lower barred labels \raisebox{.5ex}{\scalebox{0.7}{$\ytableausetup{mathmode, centerboxes,centertableaux, boxframe=normal}\begin{ytableau}*(lightblue)\overline{1}\end{ytableau}$}}, and thus appear to the right of or below lower barred numbers.

For each semitableau pair we introduce a barred sequence read in the same way as the column sequence from Algorithm~\ref{alg:column_multiplication}, i.e., bottom-to-top in each column and then columns left-to-right. Ordering barred numbers as $\overline{1}<\overline{2}$ etc., 
we show in \appref{app: equivalence of sequences} that the admissibility of the barred sequence of a semitableau pair is equivalent to the admissibility of the column sequence of the corresponding $N$-dependent ordinary semitableau.

We are now in a position to state the general algorithm for column multiplication of a Young diagram pair and a barred Young diagram:

\begin{theorem}[Multiplying a Young diagram pair by a barred diagram]
\label{alg:barred_multiplication} \ \\
    The Young diagram pairs resulting from the product $\pair{\rho}{\sigma}\otimes \overline{\mu}$ are obtained via the following procedure:
\begin{enumerate}
    \item Label each box in $\overline{\mu}$ by the column in which it appears in $\mu$, e.g., entries in column 1 of $\mu$ are labeled with $\overline{1}$ in $\overline{\mu}$.
    \item Starting from the first, i.e., left-most, column in $\overline{\mu}$, add all boxes from one column, each labeled $\overline{k}$, to a semitableau pair $\pair{\gamma}{\delta}$ obtained from a previous step, so that $\overline{\gamma}$ is extended (with solid-border boxes), or boxes from $\delta$ are removed (dashed border). In this step treat dashed boxes from previous steps as if they had been removed. Keep the resulting semitableau pair if it
        \begin{enumerate}
            \item corresponds to a Young diagram pair, i.e., when the dashed boxes are removed, the result has the shape of a Young diagram pair,
            \item has no two boxes labeled $\overline{k}$ appearing in the same row,
            \item has a barred column-sequence of numbers, read bottom-to-top and then left-to-right, that is admissible (see Definition~\ref{def:admissible}).
        \end{enumerate}
      \item After exhausting the columns of $\overline{\mu}$, for each semitableau pair, determine $N_\text{min}$ as the number of rows in that semitableau pair.
      \item Remove dashed boxes and all labels, and collect the $N$-dependent multiplicities for each diagram pair.
\end{enumerate}
\end{theorem}

Similarly to Algorithm~\ref{alg:pair_diagram_mult}, solid barred boxes can appear in  the same row as dashed ordinary boxes, if the ordinary boxes in question were removed in a step \textit{before} adding the solid barred boxes. As we add higher barred values first, e.g., the pattern 
\raisebox{0.5ex}{\scalebox{0.7}{\ytableausetup{centerboxes,centertableaux}\begin{ytableau} \overline{2} & \delboxbar{1} \end{ytableau}}}
cannot occur, but 
\raisebox{0.5ex}{\scalebox{0.7}{\ytableausetup{centerboxes,centertableaux}\begin{ytableau} \overline{1} & \delboxbar{2} \end{ytableau}}}
can 
(as \raisebox{0.5ex}{\scalebox{0.7}{\ytableausetup{centerboxes,centertableaux}\begin{ytableau} \delboxbar{2} \end{ytableau}}} should be considered removed when adding 
\raisebox{0.5ex}{\scalebox{0.7}{\ytableausetup{centerboxes,centertableaux}\begin{ytableau} \overline{1} \end{ytableau}}}).

We conclude this section by once more evaluating the product in eqs.~\eqref{eq:pair x 21bar N-dep}--\eqref{eq:pair x 21bar color barred labels}, this time using Algorithm~\ref{alg:barred_multiplication}, i.e., without ever mapping diagram pairs to $N$-dependent ordinary diagrams, 
\begin{align} \ytableausetup{mathmode, centerboxes,centertableaux, boxframe=normal}
            \scalebox{0.7}{$\ydiagram[*(white)\bullet]{0,1}*{1+1}$}
            \,\otimes \,
            \scalebox{0.7}{$\ydiagram[*(white)\bullet]{1+1,2}$} \ = \ & \ytableausetup{mathmode, centerboxes,centertableaux, boxframe=normal}\ 
            \scalebox{0.7}{$\ydiagram[*(white)\bullet]{0,1}*{1+1}$}
            \,\otimes\, 
            \scalebox{0.7}{$\begin{ytableau}
            \none & \overline{1} \\
            \overline{2} & \overline{1} \\
            \end{ytableau}$}\nonumber
            \ = \ \ytableausetup{mathmode, centerboxes,centertableaux, boxframe=normal}
    \left(\barredexamplefirstline\right)\otimes
            \scalebox{0.7}{$\begin{ytableau}
            \overline{1} \\
            \overline{1} \\
            \end{ytableau}$}\nonumber\\
 \ = \ & \ytableausetup{mathmode, centerboxes,centertableaux, boxframe=normal}\ 
\scalebox{0.7}{$\begin{ytableau}
\overline{1} & \delboxbar{2}\\
\overline{1} & \none\\
\bullet
\end{ytableau}$}\,
\oplus\, 
\scalebox{0.7}{$\begin{ytableau}
\none & \overline{1} & \delboxbar{2}\\
\overline{1} &\bullet& \none
\end{ytableau}$}
\,\oplus\,
\canceldiag{\scalebox{0.7}{$\begin{ytableau}
\none & \delboxbar{1}\\
\overline{1} \\
\overline{2} & \none\\
\bullet
\end{ytableau}$}}\,
\oplus\,
\scalebox{0.7}{$\begin{ytableau}
\none & \none & \delboxbar{1}\\
\none& \overline{2} & \none\\
\overline{1} &\bullet
\end{ytableau}$}\,
\oplus\, 
\canceldiag{\scalebox{0.7}{$\begin{ytableau}
 \none & \empty\\
 \overline{1}\\
 \overline{1} & \none\\
 \overline{2} & \none\\
 \bullet
\end{ytableau}$}}\,
\oplus\,
    \scalebox{0.7}{$\begin{ytableau}
\none & \none & \empty\\
\none & \overline{1} & \none\\
\none & \overline{2} & \none\\
\overline{1} & \bullet
\end{ytableau}$}\,
\oplus\,
\scalebox{0.7}{$\begin{ytableau}
\none & \none & \empty\\
\overline{1} & \overline{2} & \none\\
\overline{1} & \bullet
\end{ytableau}$}
\,
\nonumber\\
&  \text{\small \hspace{-5pt} $(\overline{1},\overline{1},\overline{2})$ \hspace{3pt} $(\overline{1},\overline{1},\overline{2})$ \hspace{5pt} $(\overline{2},\overline{1},\overline{1})$ \hspace{9pt} $(\overline{1},\overline{2},\overline{1})$ \hspace{12pt} $(\overline{2},\overline{1},\overline{1})$ \hspace{7pt} $(\overline{1},\overline{2},\overline{1})$
\hspace{7pt} $(\overline{1},\overline{1},\overline{2})$} \nonumber\\ \label{eq:eq:pair x 21bar final}
     \ytableausetup{mathmode, centerboxes,centertableaux, boxframe=normal}\
&\oplus\!\!\!\ytableausetup{mathmode, centerboxes,centertableaux, boxframe=normal}\
\canceldiag{\scalebox{0.7}{$\begin{ytableau}
\none & \none & \delboxbar{1}\\
\none &\overline{1} \\
 \overline{2} &\bullet & \none
\end{ytableau}$}}
\,
\oplus\, 
\scalebox{0.7}{$\begin{ytableau}
\none &\none & \none & \delboxbar{1}\\
\overline{1} & \overline{2} &\bullet & \none
\end{ytableau}$}\,
\oplus\, 
\canceldiag{\scalebox{0.7}{$\begin{ytableau}
\none & \none & \empty\\
\none & \overline{1} & \none\\
\none & \overline{1} & \none\\
\overline{2} &\bullet& \none
\end{ytableau}$}}\,
\oplus\,
\scalebox{0.7}{$\begin{ytableau}
\none & \none & \none & \empty\\
\none & \none & \overline{1} & \none\\
\overline{1} & \overline{2} &\bullet& \none
\end{ytableau}$}
\\
& \ 
 \text{\small \hspace{8pt} $(\overline{2},\overline{1},\overline{1})$ \hspace{15pt} $(\overline{1},\overline{2},\overline{1})$ \hspace{21pt} $(\overline{2},\overline{1},\overline{1})$ \hspace{17pt} $(\overline{1},\overline{2},\overline{1})$} \nonumber\\
\ = \ & \barredexampleresult \nonumber \ .
\end{align}
Here we have also displayed (and crossed out) some terms which are only ruled out by the admissibility criterion for the barred sequence. 

Notice that for $N=3$, the result of \Eqref{eq:eq:pair x 21bar final} is mapped to the same ordinary Young diagrams as the result of \Eqref{eq:pair_diagram_mult_example_N3}, even though their general-$N$ pair representations are different.

\section{Multiplying two Young diagram pairs}
\label{sec:pair multiplication}

We now combine Algorithms~\ref{alg:pair_diagram_mult}~and~\ref{alg:barred_multiplication} and derive the extra criteria necessary to reduce tensor products of any two Young diagram pairs. 

In Section~\ref{sec:algorithms_for_special_cases}, we have shown how to multiply Young diagram pairs with ordinary Young diagrams and with barred Young diagrams. A general pair is composed of a barred diagram and an ordinary Young diagram, so all that remains is to construct a criterion for combining these two special cases.

Below, we attempt to use Algorithm~\ref{alg:barred_multiplication}, followed by Algorithm~\ref{alg:pair_diagram_mult} to add boxes from a pair to another pair (here to just another Young diagram). To do so, we need to extend our notation to include the possibility of adding, say, a \raisebox{0.5ex}{\scalebox{0.7}{$\onebox{\overline{1}}$}} and then removing it with, say, a \raisebox{0.5ex}{\scalebox{0.7}{$\onebox{2}$}}. This we denote 
\raisebox{0.5ex}{\ytableausetup{centerboxes}\scalebox{0.7}{$\begin{ytableau}
\delsplitbox{1}{2}
\end{ytableau}$}}. Similarly, \raisebox{0.5ex}{\scalebox{0.7}{$\ytableausetup{centerboxes,centertableaux}\begin{ytableau}
\splitbox{1}{1}
\end{ytableau}$}} denotes an ordinary box removed by \raisebox{0.5ex}{\scalebox{0.7}{$\onebox{\overline{1}}$}} and afterwards reinserted by a \raisebox{0.5ex}{\scalebox{0.7}{$\onebox{1}$}}. 

For reference we map diagram pairs to $N$-dependent Young diagrams as in \secref{sec:pair notation}, labeling boxes derived from the barred part by negative numbers as in \secref{sec:barred multiplication}, and then apply Algorithm~\ref{alg:column_multiplication} (column multiplication). For illustration we study the minimal example {\raisebox{0.5ex}{\scalebox{0.6}{$\ydiagram{1}$}$\,\otimes\,$\scalebox{0.6}{$\ydiagram[\bullet]{0,1}*{1+1}$}}} by sequentially applying Algorithms~\ref{alg:barred_multiplication} and \ref{alg:pair_diagram_mult} (left panel below), and in parallel show column multiplication of the corresponding $N$-dependent ordinary diagrams (right panel below),
\vspace{-5pt}
\begin{equation}\label{eq:pair_mult_guess_example}
    \renewcommand{\arraystretch}{1.5} 
\begin{array}{r>{\color{black}\vrule}l>{\color{black}\vrule}l}
\begin{array}{c} 
\ytableausetup{centerboxes,centertableaux}
    \scalebox{0.7}{$\begin{ytableau}\empty\\
\end{ytableau}$}
\, \otimes\, 
\scalebox{0.7}{$\begin{ytableau}\none & \empty\\
\bullet & \none
\end{ytableau}$} 
\\
\scalebox{0.7}{$\begin{ytableau}
\none\\
\none\\
\none\\
\none\\
\none\\
\none\\
\end{ytableau}$}\\
\scalebox{0.7}{$\begin{ytableau}
\none\\
\none\\
\none\\
\none\\
\end{ytableau}$}\\
\scalebox{0.7}{$\begin{ytableau}
\none\\
\none\\
\none\\ 
\none
\end{ytableau}$}\\
\scalebox{0.7}{$\begin{ytableau}
\none\\
\none\\
\none\\
\none
\end{ytableau}$}
\end{array}
& 
\begin{array}{l} 
\ytableausetup{centerboxes}
=
\scalebox{0.7}{$\begin{ytableau}\empty\\\end{ytableau}$}
\, \otimes\, 
\scalebox{0.7}{$\begin{ytableau}\none & 1\\
\overline{1} & \none
\end{ytableau}$}
\\ [3ex]
\overset{?}{=} \ytableausetup{centerboxes}
\left(
\scalebox{0.7}{$\begin{ytableau}
 \delboxbar{1}
\end{ytableau}$}
\,\oplus\, 
\scalebox{0.7}{$\begin{ytableau}
\none & \empty\\
 \overline{1} & \none
\end{ytableau}$}
\right)\,\otimes\, 
\scalebox{0.7}{$\begin{ytableau}1
\end{ytableau}$}
\\[4ex]
\overset{?}{=}
\ytableausetup{centerboxes}
\scalebox{0.7}{$\begin{ytableau}
\splitbox{1}{1}
\end{ytableau}$}
\,\oplus\,
\scalebox{0.7}{$\begin{ytableau}
\none & \empty & 1\\
 \overline{1} &\none &\none
\end{ytableau}$}
\,\oplus\,
\scalebox{0.7}{$\begin{ytableau}
\none & \empty\\
\none & 1 \\
 \overline{1} &\none
\end{ytableau}$}
\,\oplus\,
\scalebox{0.7}{$\begin{ytableau}
\none & \empty\\
 \delsplitbox{1}{1} &\none
\end{ytableau}$}
\\[5ex]
\overset{?}{=}\ytableausetup{centerboxes}
1_1\,
\scalebox{0.7}{$\begin{ytableau}
\empty\\
\end{ytableau}$}
\,\oplus
1_2\, 
\scalebox{0.7}{$\begin{ytableau}
\none & \empty & \empty\\
\bullet &\none &\none
\end{ytableau}$}
\,\oplus
1_3\, 
\scalebox{0.7}{$\begin{ytableau}
\none & \empty\\
\none & \empty \\
\bullet &\none \\
\end{ytableau}$}
\,\oplus
1_2\,
\scalebox{0.7}{$\begin{ytableau}
\empty\\
\end{ytableau}$}
\\[5ex]
\overset{N=3}{=}
\ytableausetup{centerboxes} 
    2\ 
    \scalebox{0.7}{$\begin{ytableau}
    \empty\\
    \end{ytableau}$}
    \,\oplus\,
     \scalebox{0.7}{$\begin{ytableau}
    \empty & \empty & \\
    \empty\\
    \end{ytableau}$}
    \,\oplus\,
    \scalebox{0.7}{$\begin{ytableau}
    \empty & \empty\\
    \empty & \empty\\
    \end{ytableau}$}
\end{array} 
& 
\begin{array}{l} 
\ytableausetup{centerboxes}
    \overset{N=3}{=}\,
    \scalebox{0.7}{$\begin{ytableau}
    \none[\text{$\scriptstyle 1$}]\\
    \empty\\
    \end{ytableau}$}
    \,\otimes\,
    \scalebox{0.7}{$\begin{ytableau}
    \none[\text{\begin{rotate}{45}$\scriptstyle N\sm 1$\end{rotate}}] & \none[\text{$\scriptstyle 1$}]\\
    \sm 1 & 1\\
    \sm 1 & \none
    \end{ytableau}$} 
\\[2ex]
    = \ytableausetup{centerboxes} 
    \left(
    \scalebox{0.7}{$\begin{ytableau}
    \none[\text{$\scriptstyle N$}]\\
    \empty\\
    \sm 1\\
    \sm 1
    \end{ytableau}$}
\,\oplus\,
    \scalebox{0.7}{$\begin{ytableau}
    \none[\text{\begin{rotate}{45}$\scriptstyle N\sm 1$\end{rotate}}] &  \none[\text{$\scriptstyle 1$}]\\
    \empty & \sm 1\\
    \sm 1\\
    \end{ytableau}$}
    \right)\,\otimes\,
    \scalebox{0.7}{$\begin{ytableau}
    \none[\text{$\scriptstyle 1$}]\\
    1\\
    \end{ytableau}$}
\\[0ex]
    = \ytableausetup{centerboxes} 
    \scalebox{0.7}{$\begin{ytableau}
    \none[\text{$\scriptstyle N$}] & \none[\text{$\scriptstyle 1$}]\\
    \empty & 1\\
    \sm 1 \\
    \sm 1
    \end{ytableau}$}
    \,\oplus\,
    \scalebox{0.7}{$\begin{ytableau}
    \none[\text{\begin{rotate}{45}$\scriptstyle N\sm 1$\end{rotate}}] & \none[\text{$\scriptstyle 1$}]& \none[\text{$\scriptstyle 1$}]\\
    \empty & \sm 1 & 1\\
    \sm 1\\
    \end{ytableau}$}
    \,\oplus\,
    \scalebox{0.7}{$\begin{ytableau}
    \none[\text{\begin{rotate}{45}$\scriptstyle N\sm 1$\end{rotate}}]& \none[\text{$\scriptstyle 2$}]\\
    \empty & \sm 1\\
    \sm 1 & 1\\
    \end{ytableau}$}    
        \,\oplus\,
    \canceldiag{
    \scalebox{0.7}{$\begin{ytableau}
    \none[\text{$\scriptstyle N$}] & \none[\text{$\scriptstyle 1$}]\\
    \empty & \sm 1\\
    \sm 1 \\
    1 \\
    \end{ytableau}$}
    }
    \\[5ex]
    \ytableausetup{centerboxes}
    \scalebox{0.7}{$\begin{ytableau}
    \none
    \end{ytableau}$}
    \\[6ex]
    = \ytableausetup{centerboxes} 
    \scalebox{0.7}{$\begin{ytableau}
    \empty\\
    \end{ytableau}$}
    \,\oplus\,
    \scalebox{0.7}{$\begin{ytableau}
    \empty & \empty & \\
    \empty\\
    \end{ytableau}$}
    \,\oplus\,
    \scalebox{0.7}{$\begin{ytableau}
    \empty & \empty\\
    \empty & \empty\\
    \end{ytableau}$}\;.\\[2ex]
\end{array} \\
\end{array}
\end{equation}
In lines two and three, all terms in the left and right panels are ordered such that they are in one-to-one correspondence. In line four of the left panel we assign $N_\mathrm{min}$ values according to the number of rows in the line above. In the final line, we display the result in terms of ordinary Young diagrams for $N=3$ in both panels.

It is not surprising that there are more terms in the left panel, since the subsequent application of Algorithms~\ref{alg:barred_multiplication} and \ref{alg:pair_diagram_mult}, without further constraints, means that we effectively multiply by \raisebox{0.5ex}{$\scalebox{.6}{\ydiagram{1}}\otimes\scalebox{.6}{\ydiagram[\bullet]{1}}\,=\,\scalebox{.6}{\ydiagram[\bullet]{0,1}*{1+1}}\,\oplus\,\bullet$}.

In the right panel we discard the semitableau \[\scalebox{0.7}{$\ytableausetup{centerboxes,centertableaux}\begin{ytableau}\empty&\sm 1\\\sm 1\\1\\\end{ytableau}$}\] due to inadmissibility of its column sequence. In the left panel this term corresponds to \scalebox{0.7}{$\ytableausetup{centerboxes,centertableaux}\begin{ytableau}\none&\empty\\\delsplitbox{1}{1}&\none\end{ytableau}$} which has the labels $\overline{1},1$ in its bottom row.  Recalling that rows of a semitableau pair containing the label $\overline{1}$ correspond to rows in the respective $N$-dependent semitableau containing no label $\sm1$, we note that this semitableau pair 
must be discarded.

There is another term in the left panel, \raisebox{0.5ex}{\scalebox{0.7}{$\ytableausetup{centerboxes,centertableaux}\begin{ytableau}\splitbox{1}{1}\end{ytableau}$}}, also containing the labels $\overline{1},1$ in its bottom row. This term, however, should be kept since it corresponds to the admissible semitableau \[\scalebox{0.7}{$\ytableausetup{centerboxes,centertableaux}\begin{ytableau}\empty & 1\\\sm1\\\sm1\end{ytableau}$}\] in the right panel. We show in \appref{app:linking} that increasing $N$ inserts additional labels $\sm1$ into the column sequence of the corresponding $N$-dependent semitableau, right before the contributions from the $N$-independent columns. For $N=3$ the term \raisebox{0.5ex}{\scalebox{0.7}{$\ytableausetup{centerboxes,centertableaux}\begin{ytableau}\splitbox{1}{1}\end{ytableau}$}} with naive $N_\mathrm{min}=1$ (its number of rows) is mapped to an $N$-dependent semitableau with $3{-}1{=}2$ labels $\sm1$ preceding the label $1$ in its column sequence; for $N=2$ the same term is mapped to \raisebox{0.5ex}{\scalebox{0.7}{$\ytableausetup{centerboxes,centertableaux}\begin{ytableau}\empty&1\\\sm1\end{ytableau}$}} which still has $2{-}1{=}1$ label $\sm1$ preceding the label $1$. Consequently, we keep \raisebox{0.5ex}{\scalebox{0.7}{$\ytableausetup{centerboxes,centertableaux}\begin{ytableau}\splitbox{1}{1}\end{ytableau}$}} in the final result, but with $N_\mathrm{min}=2$. This is also the $N_\mathrm{min}$ value of the second factor of the product that we are reducing.

By the mechanism described above, pairs that are inadmissible for their naive $N_\mathrm{min}$ can be made admissible by increasing their $N_\mathrm{min}$. This is only possible if admissibly fails in the ordinary part of the pair (i.e., not if it fails in the barred part) as illustrated by the two examples above (see also Figure~\ref{fig:pair}, in \appref{app:linking}).

In example~\eqref{eq:pair_mult_guess_example}, the only constraints on top of Algorithms~\ref{alg:pair_diagram_mult} and \ref{alg:barred_multiplication} derive from the relative placement of the labels $\overline{1}$ and $1$ in a semitableau pair. This is true in general, as each semitableau pair is mapped to an $N$-dependent semitableau with a column sequence containing positive and negative labels. Admissibility of the subsequence of negative labels is guaranteed by Algorithm~\ref{alg:barred_multiplication}, and admissibility of the subsequence of positive labels is guaranteed by Algorithm~\ref{alg:pair_diagram_mult}. Thus, the total column sequence is admissible if, in addition, also every label $1$ is preceded by sufficiently many labels $\sm1$, cf.\ Definition~\ref{def:admissible}. The placement of labels $\sm1$ in the $N$-dependent semitableau is determined by the placement of the labels $\overline{1}$ in the corresponding pair. Since each row of a semitableau pair can contain at most one label $\overline{1}$ and at most one label $1$, the additional constraints can be formulated in terms of checking row-wise for the presence of each of these two kinds of labels. 

We formulate these rules in Algorithm~\ref{alg:pair_multiplication} below {in terms of another sequence, whose admissibility is checked in the usual way, see Definition~\ref{def:admissible}. We call this sequence the \textit{linking sequence} as it determines whether the barred and ordinary parts of a pair can be linked in an admissible way. In \appref{app:linking} we explain how the linking sequence achieves this.}

\begin{theorem}[Multiplying two Young diagram pairs]\label{alg:pair_multiplication} \ \\
Let $\pair{\mu}{\nu}$ and $\pair{\rho}{\sigma}$ be Young diagram pairs. The product $\pair{\mu}{\nu}\otimes\pair{\rho}{\sigma}$ is obtained by the following procedure:
\begin{enumerate}
    \item Evaluate
    $$\pair{\mu}{\nu}\otimes\overline{\rho} = \bigoplus_j \pair{\gamma_j}{\delta_j}$$ 
    using Algorithm~\ref{alg:barred_multiplication} (barred multiplication), but omitting step 4.
    \item Add the first column of $\sigma$ to the resulting semitableau pairs, following the column multiplication rules in Algorithm~\ref{alg:pair_diagram_mult}. 
    \item Go through the rows from bottom to top. Construct the \textit{linking sequence} by for each row writing
    \begin{itemize}
        \item 0 if the row contains neither {$\overline{1}$} nor {$1$}\,,
        \item 1 if the row contains \raisebox{0.75ex}{\scalebox{1.1}{$\begin{ytableau}\none[\splitbox{1}{1}]\end{ytableau}$}}\,,
        \item nothing otherwise.
    \end{itemize}
    Place a vertical bar between the entries corresponding to the barred diagram and to the ordinary Young diagram in the semitableau pair.
    \item Evaluate the linking sequence for each term. If the linking sequence is (cf.\ Definition~\ref{def:admissible})
    \begin{itemize}
        \item admissible, then keep the term without modifications,
        \item inadmissible to the left of the separating bar, then discard the term,
        \item inadmissible to the right of the separating bar, then increase the term's current $N_\text{min}$ (i.e., the number of rows of the pair including dashed boxes) by the number of additional $0$s required at the position of the bar for the sequence to become admissible. 
    \end{itemize}
    Write each semitableau pair's $N_\text{min}$ as a subscript to its multiplicity.
    \item Proceed with the column multiplication procedure, increasing a term's $N_\text{min}$ only if the number of rows grows greater than the $N_\text{min}$ for this term in the previous step.
    \item  After exhausting the columns of $\pair{\rho}{\sigma}$, remove dashed boxes and all labels, and collect the $N$-dependent multiplicities for each diagram pair.
\end{enumerate}
\end{theorem}

We illustrate the algorithm with an example, thereby also displaying, but crossing out, terms which are only ruled out due to inadmissibility of their linking sequences, 
\begin{equation}
\hspace{-25pt}
\begin{split}
\scalebox{0.8}{$\begin{ytableau}
\none & \empty\\
\bullet & \none
\end{ytableau}$}\, \otimes\,
\scalebox{0.8}{$\begin{ytableau}
\none & \empty & \empty\\
\bullet & \none \\
\bullet & \none
\end{ytableau}$}
    &= \ytableausetup{centerboxes} \left(\ \pairPairExampleLineOne\ \right)\otimes\,\scalebox{0.8}{$\ytableaushort{12}$}\\
    &= \left(\pairPairExampleLineThree 
        {
\scalebox{0.8}{$\begin{ytableau}
 \none\\
\none\\
 \none\\
\none\\
\none
\end{ytableau}$}}
    \right.
    \\
    &\text{\footnotesize \hspace{22pt} ($0\,|\, 1$) \hspace{8pt} ($0,1 \,|\,$) \hspace{20pt} ($\,|\,1$) \hspace{32pt} ($1\,|\,$) \hspace{32pt} ($0\,|\,$) \hspace{20pt} ($0\,|\,0$)}\\
    & \hspace{15pt} \,\oplus\ \left.\pairPairExampleLineFour \right)\otimes\,\scalebox{0.8}{$\ytableaushort{2}$}\\
    &\text{\footnotesize \hspace{25pt}($0,1\,|\,0$) \hspace{32pt}($\,|\,$) \hspace{33pt}($\,|\,0$) \hspace{32pt}($1\,|\,0$) \hspace{35pt}($1\,|\,0$)}\\
    \ \\
    &= \left(\ \pairPairExampleLineThreeMarked
    {
\scalebox{0.8}{$\begin{ytableau}
 \none\\
\none\\
 \none\\
\none\\
\none
\end{ytableau}$}}
    \right.\\
    & \hspace{20pt} \oplus\left.\pairPairExampleLineFourMarked 
    \right)\otimes\,\scalebox{0.8}{$\ytableaushort{2}$}\\ 
\end{split}
\end{equation}
\begin{equation}\notag
\begin{split}
    & = \pairPairExampleLineNine\\
    & \hspace{10pt}\,\oplus\,\pairPairExampleLineTen\\ \ \\
    &= \pairPairExampleLineEleven\\
    & \hspace{10pt}\,\oplus\, \pairPairExampleLineTwelve \ .
\end{split}
\end{equation}
Observe that the third term in the second line has the linking sequence $(|1)$ for which the admissibility criterion fails to the right of the bar. This term becomes admissible by increasing its $N_\mathrm{min}$ from $2$ (its number of rows) to $3$ (cf.\ Figure~\ref{fig:pair} in \appref{app:linking}). We remark that it would have been easier to evaluate 
\raisebox{0.5ex}{\scalebox{0.6}{\begin{ytableau}
\none & \empty & \empty\\
\bullet & \none \\
\bullet & \none
\end{ytableau}}$\, \otimes\,$\scalebox{0.6}{\begin{ytableau}
\none & \empty\\
\bullet & \none
\end{ytableau}}
}, 
but we used the reverse order to illustrate multiplication by a more complicated diagram pair.

\section{Conclusions}
\label{sec:conclusion}

Aiming at an analytic understanding of the QCD color structure as a function of the number of colors, we have investigated --- in an $N$-independent manner --- the reduction of tensor products of $\SU(N)$ representations. To achieve a general-$N$ decomposition, we label the representations by \textit{pairs} of Young diagrams, one ordinary Young diagram acting on fundamental representations, and one barred Young diagram acting on antifundamental representations. This, combined with a column-based version of the Littlewood--Richardson rule (Algorithm \ref{alg:column_multiplication}), paves the way for a general-$N$ version of the decomposition of the tensor product of two irreducible representations into a sum of irreducible representations. The algorithm for achieving this decomposition, which is our main result, is  presented as Algorithm~\ref{alg:pair_multiplication}. This will allow for keeping track of the $N$-dependence in representation theory-based multiplet bases or for approaches working directly with Wigner 6j coefficients.

\acknowledgments

We thank Simon Plätzer for useful comments on the manuscript. BT is supported by {\sc Villum Fonden} under project no.~29388 and the MCnet studentship funded by LPCC. \

\appendix

\section{Young diagram pairs labeling representations of {\boldmath\texorpdfstring{$\GL(N)$}{GL(N)}}}
\label{app:GLN}

We formulated all our results for representations of $\SU(N)$ having in mind applications in QCD where (anti)quarks transform under the (anti)fundamental representation of $\SU(3)$. Here we explain that our results, in particular Algorithm~\ref{alg:pair_multiplication}, when read in the right way, also hold for other classical groups, such as $\GL(N)$, $\SL(N)$ and $\U(N)$.

Let $G$ be any of the matrix Lie groups $\GL(N)$, $\SL(N)$, $\U(N)$ or $\SU(N)$, and let $V$ be a complex vector space of dimension $N$ carrying the defining representation of $G$, i.e.,\ $g\in G$ acts on elements of $V$ by matrix multiplication. Then $\overline{V}$, the dual of $V$, carries the dual representation, for which $g\in G$ acts on elements of $\overline{V}$ by multiplication with $(g^{-1})^T$.

For the unitary groups $\U(N)$ and $\SU(N)$, the inverse transpose is the same as the complex conjugate, $(g^{-1})^T=g^*$, and thus the dual representation is equivalent to the complex conjugate representation. In the main body of this work we always identified these equivalent representations and referred to them as the antifundamental representation. For $\GL(N)$ and $\SL(N)$, no such identification can be made. Hence, in this appendix we speak about the defining representation carried by $V$ and the dual representation carried by $\overline{V}$.

An ordinary Young diagram $\sigma$ with $n$ boxes labels an irreducible representation of $G$ carried by a subspace of $V^{\otimes n}$. This subspace is the image of a Young operator associated with the diagram $\sigma$, see e.g.,\ \cite{Tung:1985}. Similarly, a barred Young diagram $\overline{\rho}$ with $m$ boxes labels a subspace of $\overline{V}^{\otimes m}$, the image of a Young operator associated with $\rho$. This subspace also carries an irreducible representation of $G$. The Young diagram pair $\pair{\rho}{\sigma}$ denotes a subspace of $\overline{V}^{\otimes m}\otimes V^{\otimes n}$, obtained by acting with Young operators associated with $\rho$ and $\sigma$ on $\overline{V}^{\otimes m}$ and $V^{\otimes n}$, respectively, and keeping only those vectors of the image for which all contractions between any pair of $V$- and $\overline{V}$-factors vanish, see e.g.,\ \cite{King:1970,Tung:1985}. This subspace also carries an irreducible representation of $G$.

An ordinary Young diagram consisting of a single column of length $N$ corresponds to the one-dimensional representation for which $g\in G$ is represented by $\det g$. For $\SL(N)$ and $\SU(N)$ this is the trivial representation, but for $\GL(N)$ and $\U(N)$ it is not. In \secref{sec:notation} we introduced an operation that maps an ordinary Young diagram $\lambda$ with $n$ boxes to another ordinary Young diagram, say $\mu$, with $m$ boxes, where $n+m=cN$, and where $c$ is the number of columns of $\lambda$, cf.\ \Eqref{eq:cut out}. For $\SU(N)$ and $\SL(N)$ the representation carried by the subspace of $V^{\otimes n}$ obtained by applying a Young operator associated with $\lambda$ and the representation carried by $\overline{V}^{\otimes m}$ obtained by applying a Young operator associated with $\mu$ are equivalent, but for $\GL(N)$ and $\U(N)$ they differ by a factor of $(\det g)^c$, see e.g., \cite[Section~15.5]{Fulton:1991}. In other words, whenever we replace a column of $\ell$ barred boxes by a column of $N-\ell$ ordinary boxes we also have to multiply by $(\det g)^{-1}$, and whenever we replace a column of $N-\ell$ ordinary boxes by a column of $\ell$ barred boxes we also have to multiply by $\det g$.

In the derivations of Algorithms~\ref{alg:pair_diagram_mult}, \ref{alg:barred_multiplication}, and \ref{alg:pair_multiplication} for $\SU(N)$ we use this operation from \Eqref{eq:cut out}, but it is not part of the algorithms themselves. In the derivations we map barred diagrams and diagram pairs to ordinary Young diagrams with $N$-dependent column lengths, multiply the latter using Algorithm~\ref{alg:column_multiplication}, and finally map the resulting $N$-dependent ordinary diagrams back to $N$-independent Young diagram pairs. Repeating these steps for $\GL(N)$ produces additional factors of powers of $\det g$ in intermediate results, which, however, all cancel when mapping back to $N$-independent Young diagram pairs, since the number of columns with $N$-dependent column lengths is preserved under multiplication, cf.\ the discussion following \Eqref{eq:Nc3_adding_box}. The Algorithms~\ref{alg:pair_diagram_mult}, \ref{alg:barred_multiplication}, and \ref{alg:pair_multiplication} are thus also valid for $\GL(N)$, $\SL(N)$ and $\U(N)$ without modifications.

\section{Proof of Algorithm~\ref{alg:column_multiplication}}
\label{app:Column-wise proof}

First let us extend transposition of Young diagrams to direct sums by $\mathcal{T}(\alpha\oplus\beta)=\mathcal{T}(\alpha)\oplus\mathcal{T}(\beta)$. This allows us to also extend transposition to products by
\begin{equation}
\begin{split}
  \mathcal{T}(\lambda\otimes\mu)
  = \mathcal{T}\left(\bigoplus_\nu c_{\lambda,\mu}^\nu \, \nu \right)
  = \bigoplus_\nu c_{\lambda,\mu}^\nu \, \mathcal{T}(\nu) \, .
\end{split}
\end{equation}
Using property \eqref{eq:transposed-LR-numbers} we evaluate this further to 
\begin{equation}
\begin{split}
  \bigoplus_\nu c_{\lambda,\mu}^\nu \, \mathcal{T}(\nu)
  \overset{\eqref{eq:transposed-LR-numbers}}{=} 
  \bigoplus_\nu c_{\mathcal{T}(\lambda),\mathcal{T}(\mu)}^{\mathcal{T}(\nu)} \mathcal{T}(\nu)
  = \mathcal{T}(\lambda)\otimes\mathcal{T}(\mu)\;,
\end{split}
\end{equation}
where we have used that $\mathcal{T}$ is a bijection. 
Hence, $\mathcal{T}(\lambda\otimes\mu)=\mathcal{T}(\lambda)\otimes\mathcal{T}(\mu)$, and $\mathcal{T}^{-1}=\mathcal{T}$ implies 
\begin{equation}
    \mathcal{T}\Big(\mathcal{T}(\lambda)\otimes\mathcal{T}(\mu)\Big)=\lambda\otimes\mu \, .
\end{equation}
Consequently, evaluating both products using row multiplication (Algorithm~\ref{alg:row_multiplication}), 
we find that first transposing, then performing row multiplication, and then transposing again, gives the same result as direct row multiplication. 
This observation is illustrated in the following commutative diagram, 
\begin{equation}\label{eq:transposition_loop}
\begin{tikzcd}
\lambda \otimes \mu \arrow[rrrr, " \text{\hspace*{-2ex}row multiplication} "] \arrow[d, "\mathcal{T}"'] 
& & & & \displaystyle\bigoplus_{\nu} c_{\lambda,\mu}^\nu \, \nu  \\
\mathcal{T}(\lambda) \otimes \mathcal{T}(\mu) \arrow[rrrr, " \text{row multiplication} "]
& & & & \displaystyle\bigoplus_{\mathcal{T}(\nu)} c_{\mathcal{T}(\lambda),\mathcal{T}(\mu)}^{\mathcal{T}(\nu)} \, \mathcal{T}(\nu) \arrow[u, shift left=2ex, "\mathcal{T}"]
\end{tikzcd}  .
\end{equation}
Traversing this diagram along the lower arc 
is equivalent to column multiplication, since under transposition adding boxes row-wise is mapped to adding boxes column-wise, and the row sequence of Algorithm~\ref{alg:row_multiplication}, 2(c), is mapped to the column sequence of Algorithm \ref{alg:column_multiplication}, 2(c) (and vice versa), e.g.,\  
\begin{equation}
\ytableausetup{centerboxes,centertableaux}
\mathcal{T} \left( \scalebox{0.7}{
  $\begin{ytableau}
\empty & 1 & 1 & 1\\
 1 & 2 & 2\\
 3
\end{ytableau}$} \right)
= \scalebox{0.7}{$\begin{ytableau}
 \empty &1 & 3 \\
 1 & 2\\
1 & 2\\
1
\end{ytableau}$}\;.
\end{equation}
This concludes the proof.

\section{Proof of equivalence of admissibility of sequences}
\label{app: equivalence of sequences}

In this appendix, we prove that the barred sequence in  Algorithm~(\ref{alg:barred_multiplication}) is admissible if and only if the column
sequence of the corresponding $N$-dependent ordinary semitableau is
admissible.
To do so we discuss the order and number of occurrence of the labels $\overline{1}$ and $\overline{2}$ in barred sequences of semitableau pairs and of the labels $\sm2$ and $\sm1$ in column sequences of the corresponding $N$-dependent ordinary Young diagrams. The same arguments also apply to any pair of subsequent barred labels $\overline{k}$, $\overline{k{+}1}$ or ordinary labels $\sm(k{+}1)$, $\sm k$. For illustrating examples we rely on eqs.~\eqref{eq:pair x 21bar N-dep} and \eqref{eq:pair x 21bar color barred labels}, which we redisplay here together with the corresponding sequences,

\begin{equation}\label{eq:pair x 21bar N-dep copy}
    \begin{split}
        \ytableausetup{mathmode, centerboxes,centertableaux, boxframe=normal}
            &\scalebox{0.7}{$\ydiagram[*(white)\bullet]{0,1}*{1+1}$}
            \,\otimes \,
            \scalebox{0.7}{$\ydiagram[*(white)\bullet]{1+1,2}$}
            \ \overset{N=4}{=} \ 
            \ytableausetup{mathmode, centerboxes,centertableaux, boxframe=normal}\ 
            \scalebox{0.7}{$\begin{ytableau} 
    \none[\text{\begin{rotate}{45}$\scriptstyle N\sm 1$\end{rotate}}] & \none[\scriptstyle 1]\\
            \empty & \empty \\
            \empty \\
            \empty  \\
            \end{ytableau}$}
            \,\otimes\, 
            \scalebox{0.7}{$\begin{ytableau} 
    \none[\text{\begin{rotate}{45}$\scriptstyle N\sm 1$\end{rotate}}] & \none[\text{\begin{rotate}{45}$\scriptstyle N\sm 2$\end{rotate}}]\\
            \sm 2 &  \sm 1\\
            \sm 2 &  \sm 1\\
            \sm 2 \\
            \end{ytableau}$}
             \\[2ex]
            &\ = \  
            \ytableausetup{mathmode, centerboxes,centertableaux, boxframe=normal}
            \scalebox{0.7}{$\begin{ytableau} 
    \none[\text{\begin{rotate}{45}$\scriptstyle N\sm 1$\end{rotate}}] & \none[\text{\begin{rotate}{45}$\scriptstyle N\sm 1$\end{rotate}}] &
    \none[\text{\begin{rotate}{45}$\scriptstyle N\sm 2$\end{rotate}}] & \none[\text{$\scriptstyle 1$}]\\
            \empty & \empty & \sm 2 &  \sm 1\\
            \empty & \sm 2 &  \sm 1\\
            \empty & \sm 2 \\
            \end{ytableau}$}
            \,\oplus\,
            \scalebox{0.7}{$\begin{ytableau} 
    \none[\text{\begin{rotate}{45}$\scriptstyle N\sm 1$\end{rotate}}] & 
    \none[\text{\begin{rotate}{45}$\scriptstyle N\sm 1$\end{rotate}}] & \none[\text{\begin{rotate}{45}$\scriptstyle N\sm 1$\end{rotate}}] \\
            \empty & \empty & \sm 2 \\
            \empty & \sm 2 &  \sm 1\\
            \empty & \sm 2 &  \sm 1\\
            \end{ytableau}$}
            \,\oplus\,
            \scalebox{0.7}{$\begin{ytableau} 
    \none[\text{$\scriptstyle N$}] & 
    \none[\text{\begin{rotate}{45}$\scriptstyle N\sm 2$\end{rotate}}] & \none[\text{\begin{rotate}{45}$\scriptstyle N\sm 2$\end{rotate}}] & \none[\text{$\scriptstyle 1$}]\\
            \empty & \empty & \sm 2 &  \sm 1\\
            \empty & \sm 2 &  \sm 1\\
            \empty \\
            \sm 2 \\
            \end{ytableau}$}
            \,\oplus\,
            \scalebox{0.7}{$\begin{ytableau}
    \none[\text{$\scriptstyle N$}] & 
    \none[\text{\begin{rotate}{45}$\scriptstyle N\sm 1$\end{rotate}}] & \none[\text{\begin{rotate}{45}$\scriptstyle N\sm 3$\end{rotate}}] & \none[\text{$\scriptstyle 1$}]\\
            \empty & \empty & \sm 2 &  \sm 1\\
            \empty & \sm 2\\
            \empty  &  \sm 1\\
            \sm 2 \\
            \end{ytableau}$}
            \,\oplus\,
            \scalebox{0.7}{$\begin{ytableau} 
    \none[\text{$\scriptstyle N$}] & 
    \none[\text{\begin{rotate}{45}$\scriptstyle N\sm 1$\end{rotate}}] & \none[\text{\begin{rotate}{45}$\scriptstyle N\sm 2$\end{rotate}}] \\
            \empty & \empty & \sm 2 \\
            \empty & \sm 2 &  \sm 1\\
            \empty  &  \sm 1\\
            \sm 2 \\
            \end{ytableau}$}
            \,\oplus\,
            \scalebox{0.7}{$\begin{ytableau} 
    \none[\text{$\scriptstyle N$}] & 
    \none[\text{\begin{rotate}{45}$\scriptstyle N\sm 1$\end{rotate}}] & \none[\text{\begin{rotate}{45}$\scriptstyle N\sm 2$\end{rotate}}] \\
            \empty & \empty &  \sm 1\\
            \empty & \sm 2 &  \sm 1\\
            \empty & \sm 2\\
            \sm 2 \\
            \end{ytableau}$}
            \,\oplus\,
            \scalebox{0.7}{$\begin{ytableau} 
    \none[\text{$\scriptstyle N$}] & 
    \none[\text{$\scriptstyle N$}] & 
    \none[\text{\begin{rotate}{45}$\scriptstyle N\sm 3$\end{rotate}}] \\
            \empty & \empty &  \sm 1\\
            \empty & \sm 2 \\
            \empty & \sm 2\\
            \sm 2 &  \sm 1\\
              \end{ytableau}$}\;,
            \\ &
            \text{\tiny \qquad\quad
            $(\sm2,\sm2,\sm1$ \qquad\quad
            $(\sm2,\sm2,\sm1,$ \qquad
            $(\sm2,\sm2,\sm1,$ \qquad\quad
            $(\sm2,\sm1,\sm2,$ \qquad
            $(\sm2,\sm1,\sm2,$ \qquad
            $(\sm2,\sm2,\sm2,$ \quad\,
            $(\sm2,\sm1,\sm2,$}
          \\[-1.75ex] &
            \text{\tiny \qquad\quad\quad
            $\sm2,\sm1)$ \qquad\qquad\:
            $\sm1,\sm2)$ \qquad\quad\:
            $\sm2,\sm1)$ \qquad\qquad\:\,
            $\sm2,\sm1)$ \qquad\qquad
            $\sm1,\sm2)$ \qquad\quad\:
            $\sm1,\sm1)$ \qquad\:\:\:
            $\sm2,\sm1)$
          }\\
    \end{split}
\end{equation}
\begin{equation}\label{eq:pair x 21bar barred labels}
\ytableausetup{mathmode, centerboxes,centertableaux, boxframe=normal}
\begin{split}
\scalebox{0.7}{$\ydiagram[*(white)\bullet]{0,1}*{1+1}$}
            \,\otimes \,
    \scalebox{0.7}{$\begin{ytableau}
\none & \overline{1}\\
 \overline{2} & \overline{1}\\
\end{ytableau}$}
          \ =& \ 
\scalebox{0.7}{$\begin{ytableau}
\none & \none & \none & \empty\\
\none & \none & \overline{1} \\
\overline{1} &  \overline{2} & \bullet & \none
\end{ytableau}$}
\,\oplus\,
\scalebox{0.7}{$\begin{ytableau}
\none & \none & \none & \delboxbar{1}\\
\overline{1} &  \overline{2} & \bullet & \none
\end{ytableau}$}
\,\oplus\,
\scalebox{0.7}{$\begin{ytableau}
\none & \none & \empty \\
\overline{1} &  \overline{2}\\
\overline{1} & \bullet 
\end{ytableau}$}
\,\oplus\,
\scalebox{0.7}{$\begin{ytableau}
\none & \none & \\
\none &\overline{1}\\
\none & \overline{2}\\
\overline{1} &\bullet
\end{ytableau}$}
\,\oplus\,
\scalebox{0.7}{$\begin{ytableau}
\none & \none & \delboxbar{1}\\
\none &  \overline{2}\\
 \overline{1} &\bullet
\end{ytableau}$}
\,\oplus\,
\scalebox{0.7}{$\begin{ytableau}
\none & \overline{1} & \delboxbar{2}\\
\overline{1} &\bullet
\end{ytableau}$}
\,\oplus\,
\scalebox{0.7}{$\begin{ytableau}
\overline{1} & \delboxbar{2}\\
\overline{1}\\
\bullet
\end{ytableau}$}\;.
\\ &
\text{\small \ 
  $(\overline{1},\overline{2},\overline{1})$ \qquad\ 
  $(\overline{1},\overline{2},\overline{1})$ \qquad
  $(\overline{1},\overline{1},\overline{2})$ \quad
  $(\overline{1},\overline{2},\overline{1})$ \quad\,
  $(\overline{1},\overline{2},\overline{1})$ \quad
  $(\overline{1},\overline{1},\overline{2})$ \quad
  $(\overline{1},\overline{1},\overline{2})$ \quad
}
\end{split}
\end{equation}
We begin with two observations:
\begin{enumerate}
\item[(a)]
  If the labels $\overline{1}$ and $\overline{2}$ both appear in the same row of a semitableau pair then they have to appear as the pattern \raisebox{.5ex}{\scalebox{0.7}{$\ytableausetup{centerboxes,centertableaux}\begin{ytableau}\overline{1}&\overline{2}\end{ytableau}$}} (possibly with some or all borders dashed), as boxes labeled $\overline{2}$ are added before boxes labeled $\overline{1}$ in Algorithm~\ref{alg:barred_multiplication}, and barred boxes added first appear to the right of barred boxes added later. Similarly, if the labels $\sm2$ and $\sm1$ both appear in the same row then always as the pattern \raisebox{.5ex}{\scalebox{0.7}{$\ytableausetup{centerboxes,centertableaux}\begin{ytableau}\sm2&\sm1\end{ytableau}$}}.

  Two of these two-box patterns can be either stacked on top of each other, \raisebox{.5ex}{\scalebox{0.7}{$\ytableausetup{centerboxes,centertableaux}\begin{ytableau}\overline{1}&\overline{2}\\\overline{1}&\overline{2}\end{ytableau}$}}, in which case they contribute a subsequence $\overline{1},\overline{1},\overline{2},\overline{2}$, or they appear with the upper row shifted at least one box to the right, \raisebox{.5ex}{\scalebox{0.7}{$\ytableausetup{centerboxes,centertableaux}\begin{ytableau}\none&\overline{1}&\overline{2}\\\overline{1}&\overline{2}&\none\end{ytableau}$}}, in which case each two-box pattern contributes a subsequence $\overline{1},\overline{2}$. Similarly, for three or more such two-box patterns and correspondingly for the \raisebox{.5ex}{\scalebox{0.7}{$\ytableausetup{centerboxes,centertableaux}\begin{ytableau}\sm2&\sm1\end{ytableau}$}} patterns.

  Since all occurring subsequences are admissible, rows with any of these two-box patterns do not affect the admissibility of the barred or column sequence. We can therefore ignore rows with these two-box patterns when reading off the barred or column sequence. We refer to the sequence obtained in this way as the \textit{reduced} barred or column sequence. 
\item[(b)] All contributions of either $\sm2$ or $\sm1$ to the reduced column sequence come from rows with just  \raisebox{.5ex}{\scalebox{0.7}{$\ytableausetup{centerboxes,centertableaux}\begin{ytableau}\sm2\end{ytableau}$}} or just  \raisebox{.5ex}{\scalebox{0.7}{$\ytableausetup{centerboxes,centertableaux}\begin{ytableau}\sm1\end{ytableau}$}} (but not both), where 
  \begin{enumerate}
  \item[(i)] rows with just \raisebox{.5ex}{\scalebox{0.7}{$\ytableausetup{centerboxes,centertableaux}\begin{ytableau}\sm2\end{ytableau}$}} correspond to rows with just \raisebox{.5ex}{\scalebox{0.7}{$\ytableausetup{centerboxes,centertableaux}\begin{ytableau}\overline{1}\end{ytableau}$}} and 
  \item[(ii)] rows with just \raisebox{.5ex}{\scalebox{0.7}{$\ytableausetup{centerboxes,centertableaux}\begin{ytableau}\sm1\end{ytableau}$}} correspond to rows with just \raisebox{.5ex}{\scalebox{0.7}{$\ytableausetup{centerboxes,centertableaux}\begin{ytableau}\overline{2}\end{ytableau}$}}
  \end{enumerate}
  in the respective semitableau pair, except when a dashed box appears in the same row as a barred box of the pair. In this special case two different rows of an $N$-dependent semitableau, the lower row containing  \raisebox{.5ex}{\scalebox{0.7}{$\ytableausetup{centerboxes,centertableaux}\begin{ytableau}\sm2\end{ytableau}$}} and the upper row containing  \raisebox{.5ex}{\scalebox{0.7}{$\ytableausetup{centerboxes,centertableaux}\begin{ytableau}\sm1\end{ytableau}$}}, can correspond to a single row of a semitableau pair containing the pattern \raisebox{.5ex}{\scalebox{0.7}{$\ytableausetup{centerboxes,centertableaux}\begin{ytableau}\overline{1}&\delboxbar{2}\end{ytableau}$}}.
\end{enumerate}
For illustration, consider the reduced column sequences and the reduced barred sequences of all semitableaux and all corresponding semitableau pairs of eqs.~\eqref{eq:pair x 21bar N-dep copy} and \eqref{eq:pair x 21bar barred labels},
\begin{align}
  &(\sm2),&&(\sm2),&&(\sm2),&&(\sm2,\sm1,\sm2),&&(\sm2,\sm1,\sm2),&&(\sm2,\sm2,\sm1),&&(\sm2,\sm2,\sm1),\\
  &\ (\overline{1}),&&\ (\overline{1}),&&\ (\overline{1}),&&\;\ (\overline{1},\overline{2},\overline{1}),&&\;\ (\overline{1},\overline{2},\overline{1}),&&\qquad (\overline{1}),&&\qquad (\overline{1}).
\end{align}
For the first five terms, i.e., for semitableau pairs without the special pattern \raisebox{.5ex}{\scalebox{0.7}{$\ytableausetup{centerboxes,centertableaux}\begin{ytableau}\overline{1}&\delboxbar{2}\end{ytableau}$}} the mapping  $\overline{1}\mapsto\sm2$ and $\overline{2}\mapsto\sm1$ establishes a bijection between the reduced barred sequence of the semitableau pair and the reduced column sequence of the corresponding $N$-dependent semitableau. In the last two cases, the semitableaux contain the pattern \raisebox{.5ex}{\scalebox{0.7}{$\ytableausetup{centerboxes,centertableaux}\begin{ytableau}\overline{1}&\delboxbar{2}\end{ytableau}$}} and the reduced barred sequence corresponds to a reduced column sequence with an additional subsequence $\sm2,\sm1$, which, however, does not affect admissibility. 

It thus follows from our observations (a) and (b) that also in
general barred sequences of semitableau pairs are admissible if and
only if the column sequences of the corresponding $N$-dependent
semitableaux are admissible.

\section{The linking sequence}
\label{app:linking}

We here discuss the admissibility of semitableau pairs constructed as in \secref{sec:pair multiplication}, by subsequent application of Algorithms~\ref{alg:barred_multiplication} and \ref{alg:pair_diagram_mult}. Each semitableau pair constructed in this way corresponds to an $N$-dependent ordinary semitableau with, in general, negative and positive labels. For this semitableau's column sequence Algorithms~\ref{alg:barred_multiplication} and \ref{alg:pair_diagram_mult} ensure admissibility of the subsequence of negative and positive labels, respectively. Thus, the semitableau's total column sequence is admissible if and only if also each label $1$ is preceded by sufficiently many labels $\sm1$, cf.\ Definition~\ref{def:admissible}.

The placement of labels $\sm1$ in the $N$-dependent semitableau is determined by the placement of labels $\overline{1}$ in the corresponding semitableau pair. More precisely, a row with label $\overline{1}$ in the semitableau pair corresponds to a row without label $\sm1$ in the ordinary semitableau. As no two identical labels can appear in the same row, we have to distinguish four cases:
\begin{itemize}
\item[(1)] rows with a label $\overline{1}$ but no label $1$,
\item[(2)] rows with a label $1$ but no label $\overline{1}$,
\item[(3)] rows with labels \raisebox{0.75ex}{\scalebox{1.1}{$\begin{ytableau}\none[\splitbox{1}{1}]\end{ytableau}$}},
\item[(4)] rows with neither a label $\overline{1}$ nor a label $1$.
\end{itemize}
In the corresponding $N$-dependent ordinary semitableau, rows of type~(1) correspond to rows with neither \raisebox{.5ex}{\scalebox{0.7}{$\ytableausetup{centerboxes,centertableaux}\begin{ytableau}\sm1\end{ytableau}$}} nor \raisebox{.5ex}{\scalebox{0.7}{$\ytableausetup{centerboxes,centertableaux}\begin{ytableau}1\end{ytableau}$}}, i.e., they do not affect admissibility of the column sequence.

Rows of type~(2) correspond to rows with the pattern \raisebox{.5ex}{\scalebox{0.7}{$\ytableausetup{centerboxes,centertableaux}\begin{ytableau}\sm1&1\end{ytableau}$}}. A single such pattern does not affect the admissibility. Two of these two-box patterns can be either stacked on top of each other, \raisebox{.5ex}{\scalebox{0.7}{$\ytableausetup{centerboxes,centertableaux}\begin{ytableau}\sm1&1\\\sm1&1\end{ytableau}$}}, in which case they contribute a subsequence $\sm1,\sm1,1,1$, or they can appear with the upper row shifted at least one box to the right, \raisebox{.5ex}{\scalebox{0.7}{$\ytableausetup{centerboxes,centertableaux}\begin{ytableau}\none&\sm1&1\\\sm1&1&\none\end{ytableau}$}}, in which case each two-box pattern contributes a subsequence $\sm1,1$. Similarly, for three or more such two-box patterns. Thus, rows of type (2) also do not affect admissibility of the column sequence.

Rows of type (3) correspond to rows with just \raisebox{.5ex}{\scalebox{0.7}{$\ytableausetup{centerboxes,centertableaux}\begin{ytableau}1\end{ytableau}$}}, and rows of type~(4) correspond to rows with just \raisebox{.5ex}{\scalebox{0.7}{$\ytableausetup{centerboxes,centertableaux}\begin{ytableau}\sm1\end{ytableau}$}}. Hence, the column sequence of the $N$-dependent semitableau is admissible if rows of type~(3) are preceded by sufficiently many rows of type~(4), ordering rows bottom-to-top.

\begin{figure}[t]
  $
  \parbox{5cm}{\includegraphics[width=5cm]{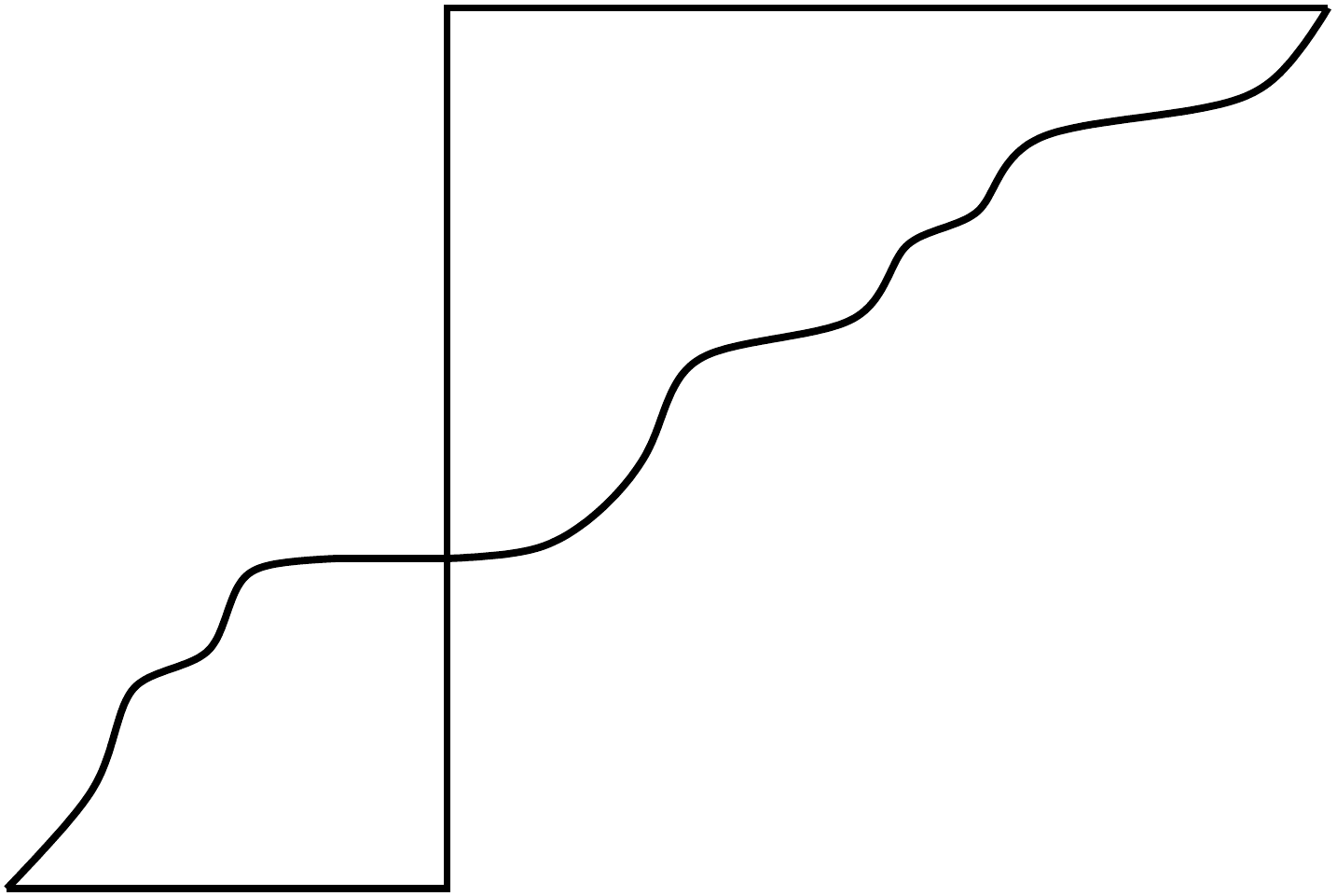}}
  = \quad \left\{ \quad \begin{matrix}
    \parbox{5cm}{\includegraphics[width=5cm]{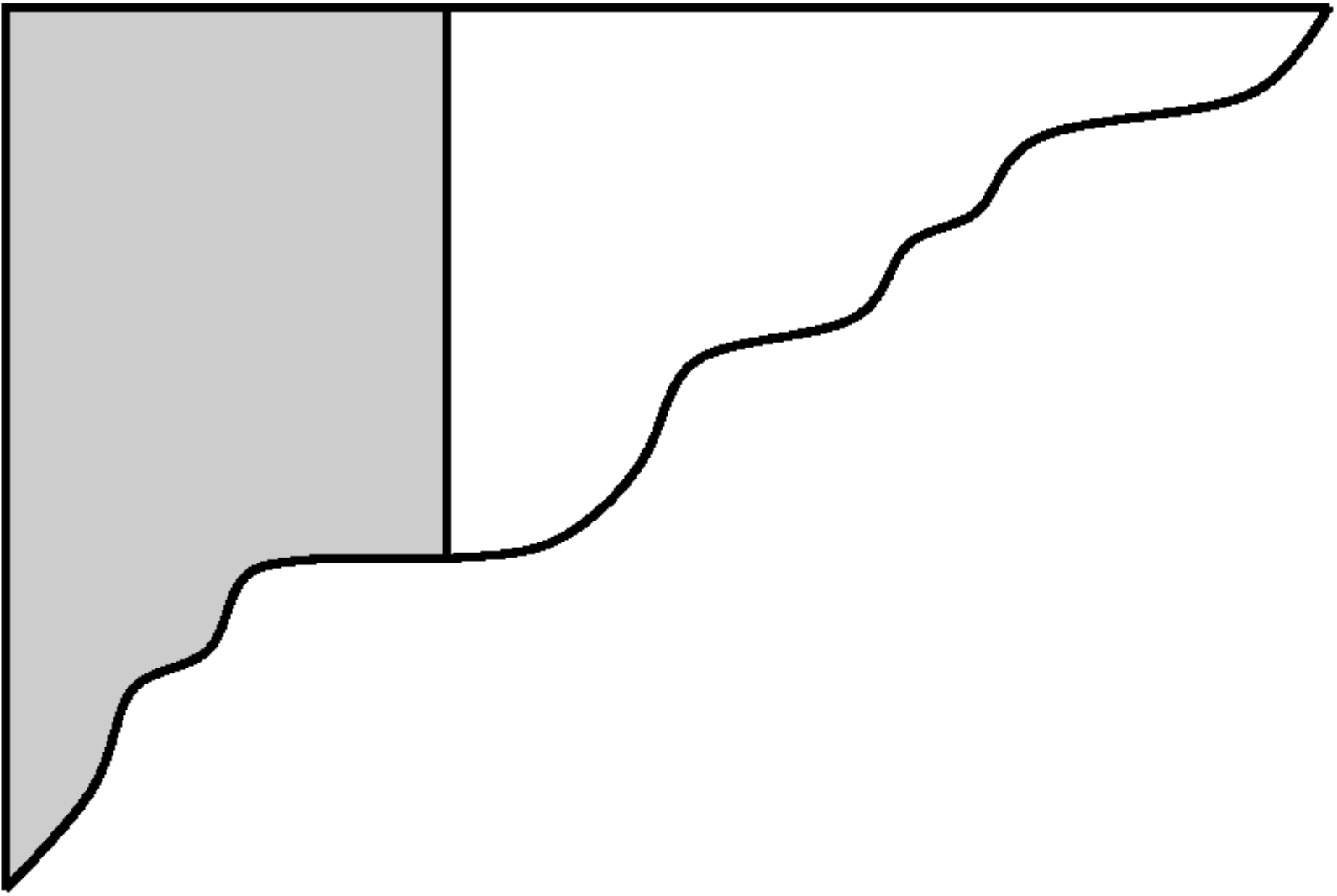}}
    & ,\ N=N_\mathrm{min}
    \\ \\
    \parbox{5cm}{\includegraphics[width=5cm]{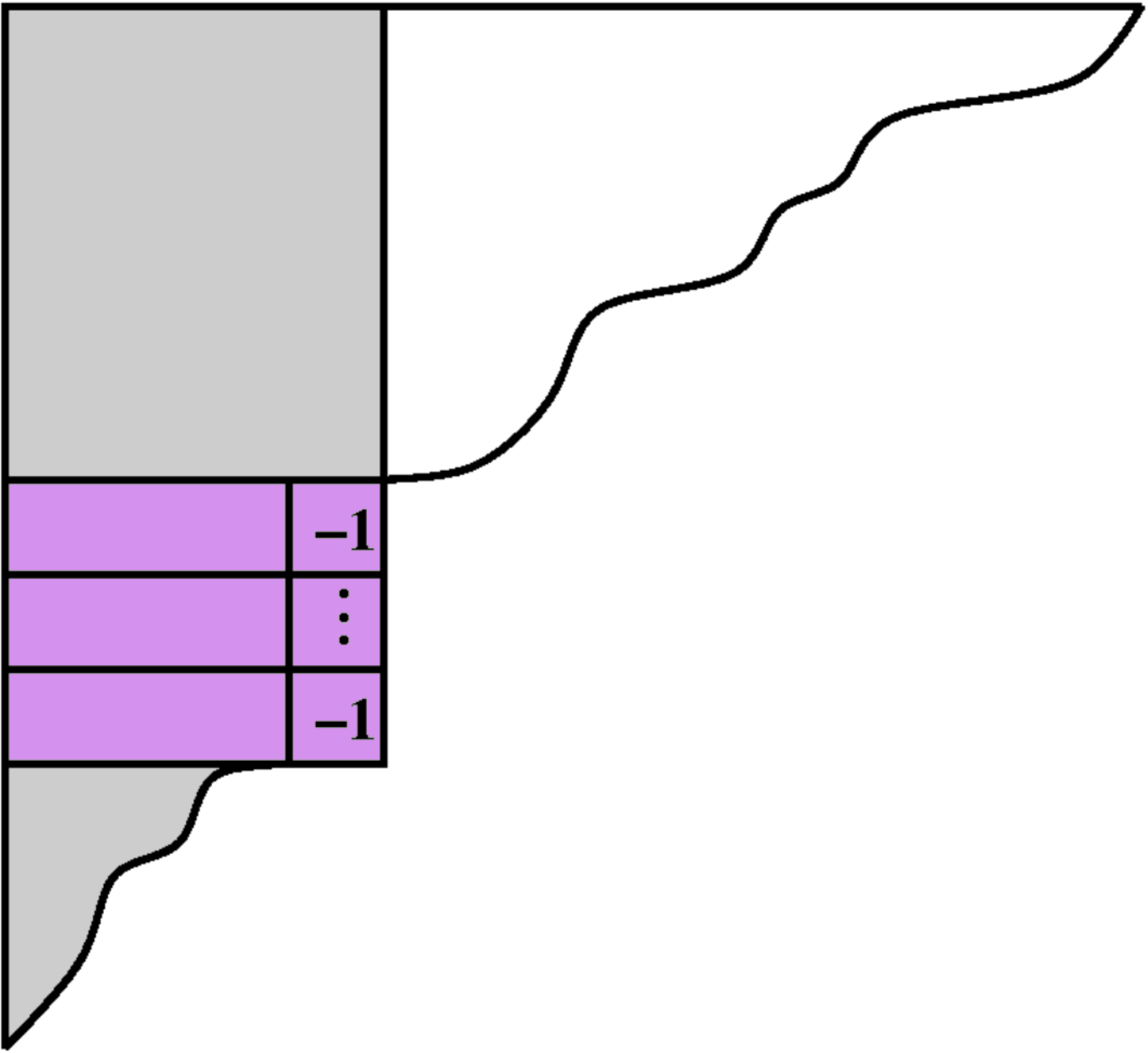}}
    & ,\ N=N_\mathrm{min}+\ell
  \end{matrix} \right.
  \begin{picture}(0,0)
    \put(-173,-67){$\left.\vphantom{\begin{matrix}A\\A\end{matrix}}\right\} \ell\text{ rows}$}
  \end{picture}
  $
  \caption{Schematic picture of an arbitrary semitableau pair mapped to an $N$-dependent ordinary semitableau for $N=N_\mathrm{min}$ and for $N=N_\mathrm{min}+\ell$, where $N_\mathrm{min}$ denotes the pair's intrinsic $N_\mathrm{min}$, i.e., its number of rows. We obtain $\ell$ additional labels $\sm1$ (from the rows shaded in purple) for $N=N_\mathrm{min}+\ell$, as compared to the case $N=N_\mathrm{min}$.
  } 
  \label{fig:pair}
\end{figure}
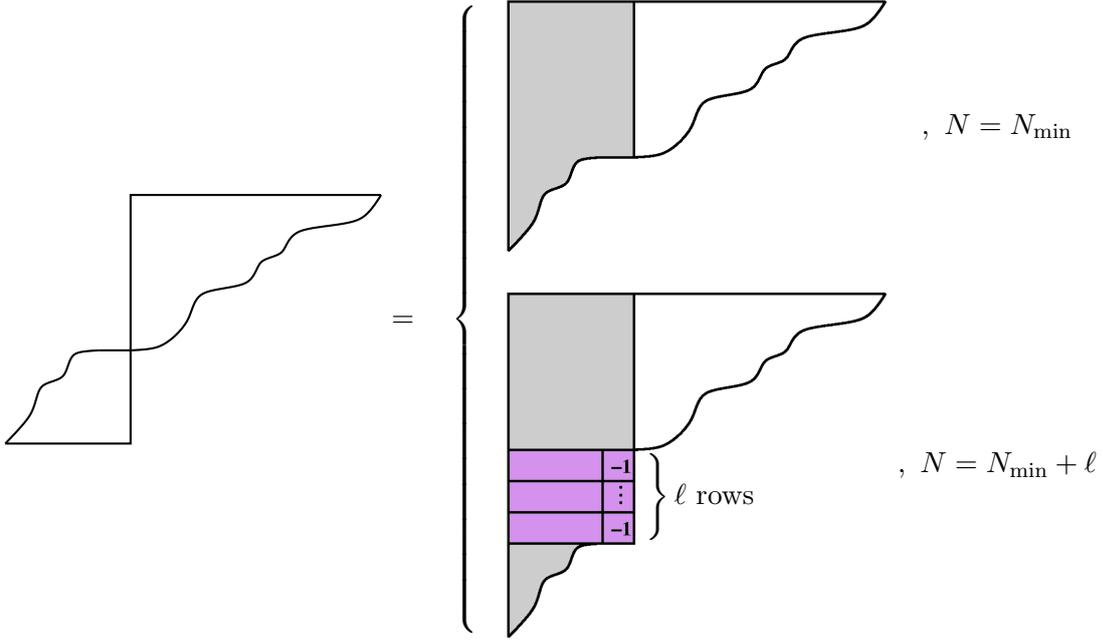

When mapping a semitableau pair with $N_\mathrm{min}$ rows to an ordinary semitableau for $N>N_\mathrm{min}$ then the latter has more rows than the pair. These additional rows each contain a label $\sm1$ (but no label $1$) as illustrated in Figure~\ref{fig:pair}. Thus, rows of type~(4) appearing above these additional rows can have labels $1$ compensated not only by rows of type~(3) but also by these additional rows. Consequently, Young diagram pairs can be inadmissible for small $N$ but become admissible for larger $N$. 

All the above considerations are encoded in the \textit{linking sequence}, which, according to Algorithm~\ref{alg:pair_multiplication}, is constructed as follows:
\begin{itemize}
\item Go the through the rows of the pair from bottom to top.
\item For each row containing \raisebox{0.75ex}{\scalebox{1.1}{$\begin{ytableau}\none[\splitbox{1}{1}]\end{ytableau}$}} write down a $1$, and for each row containing neither a $1$ nor a $\overline{1}$ write down a $0$. Ignore rows containing only one of the two types of labels.
\item When transitioning from the barred part of the diagram to the ordinary part add a vertical bar to the sequence. This is the position at which additional rows with labels $\sm1$ are created when mapping the pair to an ordinary semitableau for $N>N_\mathrm{min}$.
\end{itemize}
If the linking sequence of a pair is admissible, cf.\ Definition~\ref{def:admissible}, then the column sequence of the corresponding ordinary semitableau is admissible too. If the linking sequence fails to be admissible to the left of the bar, then the column sequence of the corresponding ordinary semitableau is inadmissible too.

If the linking sequence fails to be admissible only to the right of the bar then the column sequence of the corresponding ordinary tableau becomes admissible if $N-N_\mathrm{min}$ is at least as large as the number of additional labels $0$ required at the position of the bar for the linking sequence to become admissible.  

We illustrate the construction and use of the linking sequence with a couple of examples. In each of the following equations we display in the left panel a semitableau pair and its linking sequence. Next to each row we also write this row's contribution to the linking sequence (with a dash indicating no contribution). In the right panel we show the corresponding $N$-dependent ordinary semitableau. Note that, for reasons discussed below \Eqref{eq:Nc3_adding_box}, the number of $N$-dependent columns in the right panel is equal to the number of columns containing bullets in the left panel plus the number of distinct barred labels occurring. In the first example, 
\begin{equation}\ytableausetup{centerboxes,centertableaux}
  \begin{array}{c|c}
    \scalebox{0.8}{$\begin{ytableau}
        \none & \none & \empty & \splitbox{1}{1} & \none[\scriptstyle 1]\\
        \none & \none & \empty & 2 & \none[\scriptstyle 0]\\
        \none & \none & 1 & \none  & \none[-]\\
        \none & \delsplitbox{1}{1} & \none & \none & \none[\scriptstyle 1]\\
        \overline{1} & \overline{2} & \none & \none &\none[-]\\
        \bullet & \bullet & \none & \none &\none [\scriptstyle 0]
      \end{ytableau}$}\rightarrow (0,1\,|\, 0,1)
    \hspace*{5ex} & \hspace*{5ex} 
    \underbrace{
      \scalebox{0.8}{$\begin{ytableau}
          \none[\scriptstyle N] & \none[\scriptstyle N] & \none[\text{\begin{rotate}{45}$\scriptstyle N\sm 2$\end{rotate}}]& \none[\text{\begin{rotate}{45}$\scriptstyle N\sm 2$\end{rotate}}]& \none[\scriptstyle 3]& \none[\scriptstyle 2]\\
          \empty & \empty & \empty & \empty &\sm 2 & 1\\
          \empty & \empty & \empty & \sm 2 & \sm 1 & 2\\
          \empty & \empty & \sm 2 & \sm 1 & 1\\
          \none & \none & \none[\hspace{-1.1em}\raisebox{0.5em}{$\scriptstyle\vdots$}]\\
          \empty & \empty & \sm 2 & 1 \\
          \empty & \empty \\
          \sm 2 & \sm 1 
        \end{ytableau}$}}_{N\geq6}
  \end{array}
  \hspace*{5ex}, 
\end{equation}
the semitableau pair has six rows and its linking sequence is admissible. Consequently the column sequence of the corresponding $N$-dependent semitableau is admissible for all $N\geq6$. In the second example,
\begin{equation}\ytableausetup{centerboxes,centertableaux}
  \begin{array}{c|c}
    \scalebox{0.8}{$\begin{ytableau}
        \none & \none & \empty & \splitbox{1}{1} & \none[\scriptstyle 1]\\
        \none & \none & 1 & 2 & \none[-]\\
        \none & \delsplitbox{1}{1} & \none & \none & \none[\scriptstyle 1]\\
        \overline{1} & \overline{2} & \none & \none &\none[-]\\
        \overline{1} & \bullet & \none & \none &\none [-]
      \end{ytableau}$}\rightarrow (1\,|\, 1)
    \hspace*{5ex} & \hspace*{5ex} 
    \underbrace{\scalebox{0.8}{$\begin{ytableau}
          \none[\scriptstyle N] & \none[\text{\begin{rotate}{45}$\scriptstyle N\sm 2$\end{rotate}}]& \none[\text{\begin{rotate}{45}$\scriptstyle N\sm 2$\end{rotate}}]& \none[\scriptstyle 2]& \none[\scriptstyle 2]\\
          \empty & \empty & \empty & \sm 2 & 1\\
          \empty & \sm2 & \sm 1 & 1 & 2\\
          \none & \none[\raisebox{0.5em}{$\scriptstyle\vdots$}]\\
          *(purple)\empty & *(purple)\sm 2 & *(purple)\sm 1 \\
          \empty & \sm 2 & 1 \\
          \empty \\
          \sm 2 
        \end{ytableau}$}}_{N\geq 6}
  \end{array}
  \hspace*{5ex}, 
\end{equation}
the linking sequence fails to be admissible to the left of the bar. Despite this semitableau pair's $N_\text{min}$ is $5$, we map it to an ordinary semitableau for $N \geq 6$ to illustrate the appearance of the additional purple rows referred to in Figure~\ref{fig:pair}. Note that even if these additional rows are inserted, the semitableau pair's column sequence remains inadmissible for any $N\geq N_\text{min} = 5$. In the final example
\begin{equation}\ytableausetup{centerboxes,centertableaux}
  \begin{array}{c|c}
    \scalebox{0.8}{$\begin{ytableau}
        \none & \empty & \splitbox{1}{1} & \none[\scriptstyle 1]\\
        \none & 1 & 2 & \none[-]\\
        \delsplitbox{1}{1} & \none & \none & \none[\scriptstyle 1]\\
        \overline{1} & \none & \none &\none[-]\\
        \bullet & \none & \none &\none [\scriptstyle 0]
      \end{ytableau}$}\rightarrow (0,1\,|\, 1)
    \hspace*{5ex} & \hspace*{5ex} 
    \underbrace{\scalebox{0.8}{$\begin{ytableau}
          \none[\scriptstyle N] & \none[\text{\begin{rotate}{45}$\scriptstyle N\sm 2$\end{rotate}}]&  \none[\scriptstyle 2]& \none[\scriptstyle 2]\\
          \empty & \empty & \empty & 1\\
          \empty & \sm 1 &1 &2\\
          \empty & 1 \\
          \empty \\
          \sm 1
        \end{ytableau}$}}_{N=5}
    \qquad \text{and} \qquad  
    \underbrace{\scalebox{0.8}{$\begin{ytableau}
          \none[\scriptstyle N] & \none[\text{\begin{rotate}{45}$\scriptstyle N\sm 2$\end{rotate}}]&  \none[\scriptstyle 2]& \none[\scriptstyle 2]\\
          \empty & \empty & \empty & 1\\
          \empty & \sm 1 &1& 2\\
          \none & \none[\hspace{-1.1em}\raisebox{0.5em}{$\scriptstyle\vdots$}]\\
          *(purple)\empty & *(purple)\sm 1 \\
          \empty & 1 \\
          \empty \\
          \sm 1
        \end{ytableau}$}}_{N\geq  6}
  \end{array}
  \hspace*{5ex}, 
\end{equation}
the pair has $5$ rows and its linking sequence fails to be admissible to the right of the bar. It becomes admissible if we insert an additional $0$ at the position of the bar. Consequently this pair is mapped to a semitableau with inadmissible column sequence for $N=5$, but for $N\geq6$ it is mapped to a semitableau with admissible column sequence, due to the insertion of the additional purple row. These examples illustrate the admissibility criteria in Algorithm (\ref{alg:pair_multiplication}).

\bibliographystyle{JHEP}
\bibliography{refs}

\providecommand{\href}[2]{#2}\begingroup\raggedright\begin{thebibliography}{10}

\bibitem{Paton:1969je}
J.E.~Paton and H.-M.~Chan, \emph{Generalized {V}eneziano model with isospin},
  \href{https://doi.org/10.1016/0550-3213(69)90038-8}{\emph{Nucl. Phys. B}
  {\bfseries 10} (1969) 516}.

\bibitem{Dittner:1972hm}
P.~Dittner, \emph{{Invariant tensors in su(3). 2.}},
  \href{https://doi.org/10.1007/BF01649658}{\emph{Commun. Math. Phys.}
  {\bfseries 27} (1972) 44}.

\bibitem{Cvitanovic:1976am}
P.~Cvitanovic, \emph{{Group theory for Feynman diagrams in non-Abelian gauge
  theories}}, \href{https://doi.org/10.1103/PhysRevD.14.1536}{\emph{Phys. Rev.
  D} {\bfseries 14} (1976) 1536}.

\bibitem{Berends:1987cv}
F.A.~Berends and W.~Giele, \emph{{The Six Gluon Process as an Example of
  Weyl-Van Der Waerden Spinor Calculus}},
  \href{https://doi.org/10.1016/0550-3213(87)90604-3}{\emph{Nucl. Phys.}
  {\bfseries B294} (1987) 700}.

\bibitem{Mangano:1987xk}
M.L.~Mangano, S.J.~Parke and Z.~Xu, \emph{{Duality and Multi - Gluon
  Scattering}}, \href{https://doi.org/10.1016/0550-3213(88)90001-6}{\emph{Nucl.
  Phys.} {\bfseries B298} (1988) 653}.

\bibitem{Mangano:1988kk}
M.L.~Mangano, \emph{The color structure of gluon emission},
  \href{https://doi.org/10.1016/0550-3213(88)90453-1}{\emph{Nucl. Phys. B}
  {\bfseries 309} (1988) 461}.

\bibitem{Kosower:1988kh}
D.A.~Kosower, \emph{{Color Factorization for Fermionic Amplitudes}},
  \href{https://doi.org/10.1016/0550-3213(89)90361-1}{\emph{Nucl. Phys.}
  {\bfseries B315} (1989) 391}.

\bibitem{Nagy:2007ty}
Z.~Nagy and D.E.~Soper, \emph{Parton showers with quantum interference},
  \href{https://doi.org/10.1088/1126-6708/2007/09/114}{\emph{JHEP} {\bfseries
  09} (2007) 114} [\href{https://arxiv.org/abs/0706.0017}{{\ttfamily
  0706.0017}}].

\bibitem{Sjodahl:2009wx}
M.~Sjodahl, \emph{Color structure for soft gluon resummation -- a general
  recipe}, \href{https://doi.org/10.1088/1126-6708/2009/09/087}{\emph{JHEP}
  {\bfseries 0909} (2009) 087}
  [\href{https://arxiv.org/abs/0906.1121}{{\ttfamily 0906.1121}}].

\bibitem{Alwall:2011uj}
J.~Alwall, M.~Herquet, F.~Maltoni, O.~Mattelaer and T.~Stelzer, \emph{{MadGraph
  5 : Going Beyond}},
  \href{https://doi.org/10.1007/JHEP06(2011)128}{\emph{JHEP} {\bfseries 1106}
  (2011) 128} [\href{https://arxiv.org/abs/1106.0522}{{\ttfamily 1106.0522}}].

\bibitem{Sjodahl:2012nk}
M.~Sj{\"o}dahl, \emph{{ColorMath - A package for color summed calculations in
  SU($N_c$)}}, \href{https://doi.org/10.1140/epjc/s10052-013-2310-4}{\emph{Eur.
  Phys. J.} {\bfseries C73} (2013) 2310}
  [\href{https://arxiv.org/abs/1211.2099}{{\ttfamily 1211.2099}}].

\bibitem{Sjodahl:2014opa}
M.~Sjodahl, \emph{{ColorFull -- a C++ library for calculations in SU($N_c$)
  color space}},
  \href{https://doi.org/10.1140/epjc/s10052-015-3417-6}{\emph{Eur.Phys.J.}
  {\bfseries C75} (2015) 236}
  [\href{https://arxiv.org/abs/1412.3967}{{\ttfamily 1412.3967}}].

\bibitem{Platzer:2012np}
S.~Pl\"atzer and M.~Sjodahl, \emph{Subleading {$N_c$} improved parton showers},
  \href{https://doi.org/10.1007/JHEP07(2012)042}{\emph{JHEP} {\bfseries 07}
  (2012) 042} [\href{https://arxiv.org/abs/1201.0260}{{\ttfamily 1201.0260}}].

\bibitem{Platzer:2018pmd}
S.~Pl\"atzer, M.~Sjodahl and J.~Thor\'en, \emph{{Color matrix element
  corrections for parton showers}},
  \href{https://doi.org/10.1007/JHEP11(2018)009}{\emph{JHEP} {\bfseries 11}
  (2018) 009} [\href{https://arxiv.org/abs/1808.00332}{{\ttfamily
  1808.00332}}].

\bibitem{Frederix:2021wdv}
R.~Frederix and T.~Vitos, \emph{{The colour matrix at next-to-leading-colour
  accuracy for tree-level multi-parton processes}},
  \href{https://doi.org/10.1007/JHEP12(2021)157}{\emph{JHEP} {\bfseries 12}
  (2021) 157} [\href{https://arxiv.org/abs/2109.10377}{{\ttfamily
  2109.10377}}].

\bibitem{tHooft:1973alw}
G.~'t~Hooft, \emph{{A Planar Diagram Theory for Strong Interactions}},
  \href{https://doi.org/10.1016/0550-3213(74)90154-0}{\emph{Nucl. Phys.}
  {\bfseries B72} (1974) 461}.

\bibitem{Kanaki:2000ms}
A.~Kanaki and C.G.~Papadopoulos, \emph{{HELAC-PHEGAS: Automatic computation of
  helicity amplitudes and cross-sections}},
  \href{https://arxiv.org/abs/hep-ph/0012004}{{\ttfamily hep-ph/0012004}}.

\bibitem{Maltoni:2002mq}
F.~Maltoni, K.~Paul, T.~Stelzer and S.~Willenbrock, \emph{{Color Flow
  Decomposition of QCD Amplitudes}},
  \href{https://doi.org/10.1103/PhysRevD.67.014026}{\emph{Phys. Rev.}
  {\bfseries D67} (2003) 014026}
  [\href{https://arxiv.org/abs/hep-ph/0209271}{{\ttfamily hep-ph/0209271}}].

\bibitem{Kilian:2012pz}
W.~Kilian, T.~Ohl, J.~Reuter and C.~Speckner, \emph{{QCD in the Color-Flow
  Representation}}, \href{https://doi.org/10.1007/JHEP10(2012)022}{\emph{JHEP}
  {\bfseries 10} (2012) 022} [\href{https://arxiv.org/abs/1206.3700}{{\ttfamily
  1206.3700}}].

\bibitem{Platzer:2013fha}
S.~Pl\"atzer, \emph{{Summing Large-$N$ Towers in Colour Flow Evolution}},
  \href{https://doi.org/10.1140/epjc/s10052-014-2907-2}{\emph{Eur. Phys. J. C}
  {\bfseries 74} (2014) 2907}
  [\href{https://arxiv.org/abs/1312.2448}{{\ttfamily 1312.2448}}].

\bibitem{AngelesMartinez:2018cfz}
R.~\'Angeles~Mart\'\i{}nez, M.~De~Angelis, J.R.~Forshaw, S.~Pl\"atzer and
  M.H.~Seymour, \emph{{Soft gluon evolution and non-global logarithms}},
  \href{https://doi.org/10.1007/JHEP05(2018)044}{\emph{JHEP} {\bfseries 05}
  (2018) 044} [\href{https://arxiv.org/abs/1802.08531}{{\ttfamily
  1802.08531}}].

\bibitem{DeAngelis:2020rvq}
M.~De~Angelis, J.R.~Forshaw and S.~Pl\"atzer, \emph{{Resummation and Simulation
  of Soft Gluon Effects beyond Leading Color}},
  \href{https://doi.org/10.1103/PhysRevLett.126.112001}{\emph{Phys. Rev. Lett.}
  {\bfseries 126} (2021) 112001}
  [\href{https://arxiv.org/abs/2007.09648}{{\ttfamily 2007.09648}}].

\bibitem{Platzer:2020lbr}
S.~Pl\"atzer and I.~Ruffa, \emph{{Towards Colour Flow Evolution at Two Loops}},
  \href{https://doi.org/10.1007/JHEP06(2021)007}{\emph{JHEP} {\bfseries 06}
  (2021) 007} [\href{https://arxiv.org/abs/2012.15215}{{\ttfamily
  2012.15215}}].

\bibitem{Sjodahl:2015qoa}
M.~Sjodahl and J.~Thor{\'e}n, \emph{{Decomposing color structure into multiplet
  bases}}, \href{https://doi.org/10.1007/JHEP09(2015)055}{\emph{JHEP}
  {\bfseries 09} (2015) 055}
  [\href{https://arxiv.org/abs/1507.03814}{{\ttfamily 1507.03814}}].

\bibitem{Isaacson:2018zdi}
J.~Isaacson and S.~Prestel, \emph{{Stochastically sampling color
  configurations}},
  \href{https://doi.org/10.1103/PhysRevD.99.014021}{\emph{Phys. Rev. D}
  {\bfseries 99} (2019) 014021}
  [\href{https://arxiv.org/abs/1806.10102}{{\ttfamily 1806.10102}}].

\bibitem{Forshaw:2025bmo}
J.R.~Forshaw, S.~Pl{\"a}tzer and F.T.~Gonz{\'a}lez, \emph{{Exact colour
  evolution for jet observables}},
  \href{https://arxiv.org/abs/2502.12133}{{\ttfamily 2502.12133}}.

\bibitem{Forshaw:2025fif}
J.R.~Forshaw, S.~Pl{\"a}tzer and F.T.~Gonz{\'a}lez, \emph{{Fully Differential
  Soft Gluon Evolution at the Amplitude Level}},
  \href{https://arxiv.org/abs/2505.13183}{{\ttfamily 2505.13183}}.

\bibitem{Young:collected}
G.d.B.~Robinson, ed., \emph{The collected papers of {A}lfred {Y}oung
  (1873--1940)}, University of Toronto Press, Toronto (1977),
  \href{https://doi.org/10.3138/9781487575625}{10.3138/9781487575625}.

\bibitem{Littlewood:1934}
D.E.~Littlewood and A.R.~Richardson, \emph{Group characters and algebra},
  \href{https://doi.org/10.1098/rsta.1934.0015}{\emph{Philos. Trans. R. Soc. A}
  {\bfseries 233} (1934) 99}.

\bibitem{King:1970}
R.C.~King, \emph{Generalized {Y}oung tableaux and the general linear group},
  \href{https://doi.org/10.1063/1.1665059}{\emph{J. Math. Phys.} {\bfseries 11}
  (1970) 280}.

\bibitem{Sjodahl:2018cca}
M.~Sjodahl and J.~Thor\'en, \emph{{QCD multiplet bases with arbitrary parton
  ordering}}, \href{https://doi.org/10.1007/JHEP11(2018)198}{\emph{JHEP}
  {\bfseries 11} (2018) 198}
  [\href{https://arxiv.org/abs/1809.05002}{{\ttfamily 1809.05002}}].

\bibitem{MacFarlane:1968vc}
A.~Macfarlane, A.~Sudbery and P.~Weisz, \emph{On {G}ell-{M}ann's
  $\lambda$-matrices, $d$- and $f$-tensors, octets, and parametrizations of
  {SU(3)}}, {\emph{Commun. Math. Phys.} {\bfseries 11} (1968) 77}.

\bibitem{Butera:1979na}
P.~Butera, G.M.~Cicuta and M.~Enriotti, \emph{{Group Weight and Vanishing
  Graphs}}, \href{https://doi.org/10.1103/PhysRevD.21.972}{\emph{Phys. Rev.}
  {\bfseries D21} (1980) 972}.

\bibitem{Kyrieleis:2005dt}
A.~Kyrieleis and M.H.~Seymour, \emph{The colour evolution of the process $qq
  \to qqg$}, {\emph{JHEP} {\bfseries 01} (2006) 085}
  [\href{https://arxiv.org/abs/hep-ph/0510089}{{\ttfamily hep-ph/0510089}}].

\bibitem{Dokshitzer:2005ig}
Y.L.~Dokshitzer and G.~Marchesini, \emph{Soft gluons at large angles in hadron
  collisions}, {\emph{JHEP} {\bfseries 01} (2006) 007}
  [\href{https://arxiv.org/abs/hep-ph/0509078}{{\ttfamily hep-ph/0509078}}].

\bibitem{Sjodahl:2008fz}
M.~Sjodahl, \emph{Color evolution of 2 $\to$ 3 processes},
  \href{https://doi.org/10.1088/1126-6708/2008/12/083}{\emph{JHEP} {\bfseries
  12} (2008) 083} [\href{https://arxiv.org/abs/0807.0555}{{\ttfamily
  0807.0555}}].

\bibitem{Beneke:2009rj}
M.~Beneke, P.~Falgari and C.~Schwinn, \emph{{Soft radiation in heavy-particle
  pair production: All-order colour structure and two-loop anomalous
  dimension}},
  \href{https://doi.org/10.1016/j.nuclphysb.2009.11.004}{\emph{Nucl. Phys. B}
  {\bfseries 828} (2010) 69} [\href{https://arxiv.org/abs/0907.1443}{{\ttfamily
  0907.1443}}].

\bibitem{Keppeler:2012ih}
S.~Keppeler and M.~Sjodahl, \emph{{Orthogonal multiplet bases in {$\SU(N_c)$}
  color space}}, \href{https://doi.org/10.1007/JHEP09(2012)124}{\emph{JHEP}
  {\bfseries 09} (2012) 124} [\href{https://arxiv.org/abs/1207.0609}{{\ttfamily
  1207.0609}}].

\bibitem{Du:2015apa}
Y.-J.~Du, M.~Sj{\"o}dahl and J.~Thor{\'e}n, \emph{{Recursion in multiplet bases
  for tree-level MHV gluon amplitudes}},
  \href{https://doi.org/10.1007/JHEP05(2015)119}{\emph{JHEP} {\bfseries 05}
  (2015) 119} [\href{https://arxiv.org/abs/1503.00530}{{\ttfamily
  1503.00530}}].

\bibitem{Keppeler:2013yla}
S.~Keppeler and M.~Sj\"odahl, \emph{{Hermitian Young Operators}},
  \href{https://doi.org/10.1063/1.4865177}{\emph{J. Math. Phys.} {\bfseries 55}
  (2014) 021702} [\href{https://arxiv.org/abs/1307.6147}{{\ttfamily
  1307.6147}}].

\bibitem{Alcock-Zeilinger:2016bss}
J.~Alcock-Zeilinger and H.~Weigert, \emph{Simplification rules for birdtrack
  operators}, \href{https://doi.org/10.1063/1.4983477}{\emph{J. Math. Phys.}
  {\bfseries 58} (2017) 051701}
  [\href{https://arxiv.org/abs/1610.08801}{{\ttfamily 1610.08801}}].

\bibitem{Alcock-Zeilinger:2016sxc}
J.~Alcock-Zeilinger and H.~Weigert, \emph{{Compact Hermitian Young Projection
  Operators}}, \href{https://doi.org/10.1063/1.4983478}{\emph{J. Math. Phys.}
  {\bfseries 58} (2017) 051702}
  [\href{https://arxiv.org/abs/1610.10088}{{\ttfamily 1610.10088}}].

\bibitem{Alcock-Zeilinger:2016cva}
J.~Alcock-Zeilinger and H.~Weigert, \emph{{Transition Operators}},
  \href{https://doi.org/10.1063/1.4983479}{\emph{J. Math. Phys.} {\bfseries 58}
  (2017) 051703} [\href{https://arxiv.org/abs/1610.08802}{{\ttfamily
  1610.08802}}].

\bibitem{Ohl:2024}
T.~Ohl, \emph{{B}irdtracks of exotic {$\SU(N)$} color structures},
  \href{https://doi.org/10.1007/JHEP06(2024)203}{\emph{JHEP} {\bfseries 2024}
  (2024) 203} [\href{https://arxiv.org/abs/2403.04685}{{\ttfamily
  2403.04685}}].

\bibitem{Sjodahl:2024fqn}
M.~Sjodahl, \emph{{Orthogonal color bases for exotic representations}},
  \href{https://doi.org/10.1016/j.physletb.2025.139715}{\emph{Phys. Lett. B}
  {\bfseries 868} (2025) 139715}
  [\href{https://arxiv.org/abs/2412.07390}{{\ttfamily 2412.07390}}].

\bibitem{Alcock-Zeilinger:2022hrk}
J.~Alcock-Zeilinger, S.~Keppeler, S.~Pl\"atzer and M.~Sjodahl, \emph{{Wigner 6j
  symbols for {$\SU(N)$}: Symbols with at least two quark-lines}},
  \href{https://doi.org/10.1063/5.0131538}{\emph{J. Math. Phys.} {\bfseries 64}
  (2023) 023504} [\href{https://arxiv.org/abs/2209.15013}{{\ttfamily
  2209.15013}}].

\bibitem{Keppeler:2023msu}
S.~Keppeler, S.~Pl\"atzer and M.~Sjodahl, \emph{{Wigner 6j symbols with gluon
  lines: completing the set of 6j symbols required for color decomposition}},
  \href{https://doi.org/10.1007/JHEP05(2024)051}{\emph{JHEP} {\bfseries 05}
  (2024) 051} [\href{https://arxiv.org/abs/2312.16688}{{\ttfamily
  2312.16688}}].

\bibitem{Tung:1985}
W.-K.~Tung, \emph{{G}roup {T}heory in {P}hysics}, World Scientific Publishing
  Co., Philadelphia, PA (1985).

\bibitem{Cvitanovic:2008}
P.~Cvitanovi{\'c}, \emph{{G}roup {T}heory: {B}irdtracks, {L}ie's, and
  {E}xceptional {G}roups}, Princeton University Press, Princeton, NJ (2008).

\bibitem{Littlewood:1950}
D.E.~Littlewood, \emph{{T}he {T}heory of {G}roup {C}haracters and {M}atrix
  {R}epresentations of {G}roups}, Oxford University Press, London (1950).

\bibitem{Hamermesh:1962}
M.~Hamermesh, \emph{{G}roup {T}heory and its {A}pplication to {P}hysical
  {P}roblems}, Addison-Wesley, Reading, MA (1962).

\bibitem{Lichtenberg:1978}
D.B.~Lichtenberg, \emph{Unitary {S}ymmetry and {E}lementary {P}articles},
  Academic Press, New York, San Francisco, London, 2nd~ed. (1978).

\bibitem{Jones:1990}
H.F.~Jones, \emph{Groups, {R}epresentations and {P}hysics}, Adam Hilger, Ltd.,
  Bristol (1990),
  \href{https://doi.org/10.1887/0750305045}{10.1887/0750305045}.

\bibitem{Fulton:1997}
W.~Fulton, \emph{{Y}oung {T}ableaux}, Cambridge University Press, Cambridge
  (1997).

\bibitem{Sagan:2000}
B.E.~Sagan, \emph{{T}he {S}ymmetric {G}roup - {R}epresentations,
  {C}ombinatorical {A}lgorithms, and {S}ymmetric {F}unctions}, Springer, New
  York, 2nd~ed. (2000),
  \href{https://doi.org/10.1007/978-1-4757-6804-6}{10.1007/978-1-4757-6804-6}.

\bibitem{King_1970}
R.C.~King, \emph{The dimensions of irreducible representations of linear
  groups}, \href{https://doi.org/10.4153/CJM-1970-052-6}{\emph{Canadian Journal
  of Mathematics} {\bfseries 22} (1970) 436–448}.

\bibitem{GREBlack_1983}
G.R.E.~Black, R.C.~King and B.G.~Wybourne, \emph{Kronecker products for compact
  semisimple lie groups},
  \href{https://doi.org/10.1088/0305-4470/16/8/006}{\emph{Journal of Physics A:
  Mathematical and General} {\bfseries 16} (1983) 1555}.

\bibitem{CJCummins_1987}
C.J.~Cummins and R.C.~King, \emph{{Composite Young diagrams, supercharacters of
  U(M/N) and modification rules}},
  \href{https://doi.org/10.1088/0305-4470/20/11/018}{\emph{Journal of Physics
  A: Mathematical and General} {\bfseries 20} (1987) 3121}.

\bibitem{Fulton:1991}
W.~Fulton and J.~Harris, \emph{Representation {T}heory. {A} first course},
  Springer-Verlag, New York (1991),
  \href{https://doi.org/10.1007/978-1-4612-0979-9}{10.1007/978-1-4612-0979-9}.

\end{thebibliography}\endgroup

\end{document}